\documentclass[a4paper,10pt,journal,twoside,compsoc]{IEEEtran}
\usepackage[numbers,sort&compress,square]{natbib}

\ifCLASSINFOpdf
   \usepackage[pdftex]{graphicx}

\else
 \fi

\usepackage{array}

\usepackage{amsmath,amssymb, amsbsy,bm}
\usepackage{commath}
\usepackage{xcolor}
\usepackage{booktabs}
\usepackage[caption=false,labelformat=simple]{subfig}

\usepackage{url}
\usepackage[normalem]{ulem}
\usepackage{pinlabel}
\usepackage[breaklinks, hidelinks,colorlinks=false]{hyperref}
\usepackage{enumerate}

\newcolumntype{L}[1]{>{\arraybackslash}p{#1}}
\newcolumntype{X}[1]{>{\raggedright\let\newline\\\arraybackslash\hspace{0pt}}m{#1}}

\begin{document}
\title{graphTPP: A multivariate based method for interactive graph layout and analysis}
\author{Helen Gibson and Paul Vickers%
\thanks{H. Gibson is with the Department of Computing, Sheffield Hallam University, Sheffield, UK, e-mail: h.gibson@shu.ac.uk.}%
\thanks{P. Vickers is with the Department of Computer and Information Sciences, Northumbria University, Newcastle upon Tyne, UK, e-mail: paul.vickers@northumbria.ac.uk.}}
\markboth{Pre-print}%
{Gibson \& Vickers: graphTPP --- multivariate interactive graph layout}

\IEEEtitleabstractindextext{%
\begin{abstract}
Graph layout is the process of creating a visual representation of a graph through a node-link diagram. Node-attribute graphs have additional data stored on the nodes which describe certain properties of the nodes called attributes. Typical force-directed representations often produce hairball-like structures that neither aid in understanding the graph's topology nor the relationship to its attributes. The aim of this research was to investigate the use of node-attributes for graph layout in order to improve the analysis process and to give further insight into the graph over purely topological layouts. In this article we present graphTPP, a graph based extension to targeted projection pursuit (TPP) --- an interactive, linear, dimension reduction technique --- as a method for graph layout and subsequent further analysis. TPP allows users to control the projection and is optimised for clustering. Three case studies were conducted in the areas of influence graphs, network security, and citation networks. In each case graphTPP was shown to outperform standard force-directed techniques and even other dimension reduction methods in terms of clarity of clustered structure in the layout, the association between the structure and the attributes and the insights elicited in each domain area. 
\end{abstract}
\begin{IEEEkeywords}
Adjacency matrix, Node attributes, Graph visualization, Targeted projection pursuit
\end{IEEEkeywords}}
\maketitle

\section{Introduction}

A graph describes a set of relationships between  entities. Graph layout, the process of creating a visual representation of those relationships as a node-link diagram, is well explored in both the graph drawing and information visualisation communities. Laying out a graph, as opposed to analysing its statistical properties, is considered a vital step in the process of understanding its structure, eliciting insights, and identifying interesting or unexpected patterns and outliers~\cite{Bezerianos_2010,Henry_2006}.

The two communities approach the problem from different perspectives: graph drawing focuses on adhering to aesthetic criteria (e.g., reducing edges crossings, uniform node distribution and edge lengths) while the goal in graph visualisation is context-dependent, that is, to understand the data better through visualisation. Nevertheless, the goals of the two communities are not mutually exclusive.  

Due to this context-dependent nature of graph visualisation, the data we want to display extends beyond relationships; it encompasses node or edge attributes, clusters and groups, and multiple node and edge types. These attributes are often significantly related to a graph's structure~\cite{Newman_2003, Kim_2010, Kim_2011, Kim_2012} thus incorporating them into the visualisation enriches the analysis; the question is how should they be represented? A  layout with attributes only represented through colour, shape and size may lack interaction, specifically interactive data exploration, for visual analysis and sense-making~\cite{Endert_2011}. A layout that emerges from interactive exploration should be more meaningful to the user because they have been immersed inside the data context. They gain insights through layout creation, can use these insights to form new hypotheses, follow new ideas and ask  questions of the data~\cite{Elmqvist_2011}, explore `what-if' scenarios, and begin to infer correlations between the graph's layout,  relationships, attributes and clusterings. 

Few methods for drawing graphs allow for both the presentation and the analysis of the graph concurrently allowing one to guide the other. Thus a visual analytics application for graph visualisation could elicit more meaningful insights from the graph and the resulting layout than  other graph visualisation methods. To this end, we present a layout solution, graphTPP, which tackles both the multivariate aspect of graph layout and the interactivity that is often missing from the visual analysis process. Three case studies are explored which lay the foundation for identifying gaps in the area of multivariate graph layout and enable us to establish some grounds for future work. 

\section{Related Work}\label{sec:rl}

A graph is defined as a set of nodes and associated edges that describe a relationship between two nodes. Graph visualisations can help make sense of the graph's structure; however, how the graph is drawn has a significant impact on how it is understood. For example, due to the Gestalt principle of proximity, a relationship is inferred between nodes placed close together~\cite{McGrath_1996} influencing the user's perception of the graph. Therefore, finding a layout which emphasises relationships, without being misleading, is crucial even if further interaction, filtering and analysis may be necessary.

An extensive review of the main approaches to graph layout and visualisation can be found in Gibson, Faith and Vickers~\cite{Gibson_2013} so to avoid unnecessary repetition, the main points are summarised below along with pointers to more recent relevant research and the reader is referred to the earlier source for a more detailed treatment. 

The purpose of network visualisation is to aid the analysis and understanding of a graph. Analysing a graph's structure may involve statistical node and graph metric analysis~\cite{Wasserman_1994}, or a visual representation. Visualisation can help with ``detecting, understanding and identifying unexpected patterns''~\cite{Bezerianos_2010} in social networks and, in fact, this could be applied to all graphs while a layout has even been said to be ``necessary to find insight''~\cite{Perer_2008} by allowing users to see relationships, such as patterns and outliers, not apparent through metrics-based analysis alone.

The layout should also support the user's exploration and analysis and this will be partly governed the the view they have. Lima~\cite{Lima_2011} proposes three fundamental views for network visualisation. (1) macro: for overviews and patterns emphasising the network's structure; (2) relationship: the main level for analysis focusing on edge relationships and the relationships implied by node proximities; and (3) micro: for directing attention onto individual nodes to identify their attributes and characteristics. A good overview, displayed by the  macro and relationship views, has been described as a challenge for larger graphs~\cite{Lee_2006} and should facilitate tasks such as detecting patterns, clusters and outliers. 

Drawing a graph can enable both hypothesis generation and confirmation from the data. Much of the research into what can be learnt from studying the properties of a graph either concentrates on network analysis statistics or the study of the topological, static display of the network~\cite{Bezerianos_2010}. Layouts that stem from interaction with the network, integration of node-attributes and node metrics encourage further exploration and understanding of the network. The more information known about the graph the greater the emphasis is on good layout algorithms to convey that information in a way which is informative, accessible and comprehensible and that will instigate interaction and engagement with the graph by the user. 

The next sections present a brief review of existing layout methods. 

\subsection{Force-directed Layout} 
Graph layouts are often based on the force-directed paradigm of modelling a graph as a physical system where nodes are attracted and repelled according to some force that optimises some aesthetic criteria. Force-directed algorithms are common and often based on Eades' spring-embedded layout~\cite{Eades_1984}. Nodes are modelled as steel rings and edges as springs. From an initial random configuration the system is released and reaches  a stable state where the force on each node is zero; the optimal layout. Connected nodes attract, while other nodes repel resulting in a layout with uniform edge lengths and symmetry. Eades stressed this layout was suitable only for graphs with fewer than 50 nodes and underlying structures such as grids, trees and sparse graphs meaning that the algorithm may produce poor layouts for larger graphs, a common problem in other force-directed techniques.

Energy-based layouts treat layout as an optimisation problem with an energy function that  encodes the desired properties of the graph. The global minimum is equivalent to the optimal layout. The main energy-based layouts are Kamada and Kawai's~\cite{KK_1989}, where Euclidean distance is used to approximate graph-theoretic distance (the shortest path between nodes) and minimises the sum of squares difference between them, and simulated annealing~\cite{Davidson_1996} which mimics the annealing process of cooling a liquid slowly to form a minimal energy crystalline structure whereby a cost function encoding various layout aesthetics is gradually minimised. 

Two other notable energy-based techniques are Noack's LinLog layouts~\cite{Noack_2003, Noack_2007} which have linear attractive forces but logarithmic repulsive forces, and ForceAtlas~\cite{Jacomy_2011} from Gephi~\cite{Bastian_2009} which initially prioritises speed over precision using linear attractive and repulsive forces.  

\subsection{Multivariate Graph Layout}\label{sec:mgl}

Graphs with node attributes are termed multivariate graphs. Graphs with these properties are common; for example, personal data in social networks or gene expression data in protein interaction networks~\cite{Freeman_2005, Gehlenborg_2010, Shannon_2003, Yildirim_2007}. Typically this data is encoded as retinal variables such as the colour (and colour gradients), shape, and size of the nodes or replacing them with glyphs. However, attributes can also be used to influence layout (i.e., node position). Here we define an attribute as~\cite{Gibson_2013}:

\begin{enumerate}[(a)]
\item a piece of data  about a node (or edge) that already exists;
\item a derived item of data about a node such as a computed centrality metric or a cluster generated from an algorithm;
\item a user-defined restriction that constrains a node's position. 
\end{enumerate}

Further, while temporal data (where a node's state is linked to a particular date and time) is also a node attribute we consider it a special case which has its own set of layout algorithms and methods.  

Studies in network science support the idea that correlations between a graph's structure and its attributes exist. Newman~\cite{Newman_2003} described mixing patterns in social networks to be assortative --- the tendency for nodes to be connected to similar nodes. However, technological and biological networks were more likely to be disassortative by node degree. Heer and Perer~\cite{Heer_2011} expect these correlations to increase the potential insight that can be gained through a visualisation. 

Attributes have been used to show structural patterns in layout since the 1930s~\cite{Freeman_2000} including for pass networks~\cite{Moreno_1934},  sociometric status for social networks~\cite{Lundberg_1938, Northway_1940}, and preventative measures for HIV using node centrality~\cite{Brandes_2004, Brandes_2003b}.

There are three main ways we can use attributes to influence the layout of the graph:

\begin{enumerate}
	\item impose restrictions on the placement of nodes, e.g., inside a specific area,
	\item use membership of a group to clusters nodes, or 
	\item directly map an attribute (or attributes) onto coordinates in the layout space (e.g., $x$ and $y$ in a Cartesian system). 
\end{enumerate}
Below, we discuss some of the layouts developed specifically for these cases. 

\subsubsection{Constraint Based Layout}\label{sec:cbl}
Constraint-based techniques impose user-defined placement criteria on all or a selection of the nodes in addition to a layout algorithm (often force-directed~\cite{DiBattista_1999}). Constraints includes fixed node position, group separation, or specific layouts for a sub-graphs.  

Sugiyama \emph{et al.}'s~\cite{Sugiyama_1981} approach for hierarchical structures is a well-known constraint-based layout resulting in a layered style where first vertical, then horizontal positions are assigned to reduce edge crossings. Both social status~\cite{Brandes_2001} and clusters~\cite{Brandes_2004} have been used to assign nodes to a layer. 

Other constraint based approaches include He and Marriott's~\cite{He_1998} which aims to preserve the users' mental map, Shneiderman and Aris's Semantic Substrates~\cite{Aris_2007}, and those by Dwyer~\cite{Dwyer_2005, Dwyer_2006a}. DiG-CoLa~\cite{Dwyer_2005, Dwyer_2006a} is a derivative of Sugiyama's method that additionally optimises on aesthetic principles for directed graphs and adds orthogonal ordering constraints that restrict node position in relation to other nodes~\cite{Dwyer_2006c}. These methods have recently been encapsulated in an updated JavaScript library called Web-CoLa.\footnote{\url{http://marvl.infotech.monash.edu.au/webcola/}}

\subsubsection{Clustering-based Layouts}\label{sec:clubl}

Revealing clustering through layout is an efficient way of communicating a graph's structure. Users will often  neglect the aesthetic of edge crossing in favour of clustering~\cite{van_Ham_2008}. Both force and placement restriction methods use clustering. Force-based methods include those by Noack~\cite{Noack_2003, Noack_2007}, OpenOrd~\cite{Martin_2011}, and those which add a virtual node for each cluster with virtual edges to members of that cluster followed by a force-directed algorithm~\cite{Garcia_2007,Huang_1998}. GraphScape clusters nodes based on attribute similarity followed by a modified spring-embedder~\cite{Xu_2007}.

Methods also exist that explicitly show clusters such as the Group-in-a-Box layout~\cite{Rodrigues_2011} that places each cluster into the rectangle of a treemap which are then laid out individually within the rectangle. Treemaps have also been utilised for layout \cite{Fekete_2003,Muelder_2008} where the graph is decomposed into a tree structure and the resulting layout is in treemap form with overlaid edges. Space-filling curve methods that position nodes along the curve according to some computed ordering have also been explored, scaling to graphs with over one million nodes~\cite{Muelder_2008b}. For an in depth review of visualising group structures in graphs see Vehlow, Beck and Weiskopf's state-of-the-art report~\cite{vehlow2015}. 

\subsubsection{Mapping attributes to two-dimensions}\label{sec:drbl}

Attributes that already represent a position such as geo-coordinates~\cite{Becker_1995, Leydesdorff_2010} or those on a sports field~\cite{Dharmawirya_2010, Moreno_1934} are often mapped to two-dimensional space. Directly mapping attributes to Cartesian coordinates in 2D space is also possible. The aggregated graph layout, PivotGraph~\cite{Wattenberg_2006}, based on the idea of pivot tables, produces a grid-based graph of two categorical attributes with node sizes representing attribute occurrence. Exploration of the relationships between attributes in the graph takes place through collapsing and expanding the attribute nodes; however, it also can obscure the topology of the graph by making graphs incorrectly appear connected or cyclic. GraphDice~\cite{Bezerianos_2010} supports exploratory graph layout and allows users to lay out the graph using two attributes each on one axis. An overview of the correlation between different attributes is also shown enabling the user to easily move between them. 

\subsection{Existing Dimension Reduction based layouts}\label{sec:dr}

Dimension reduction techniques have been applied for both topological layout and multivariate layout. Dimension reduction takes data expressed in high-dimensional space and projects it onto a lower-dimensional space. The challenge is to capture the high-dimensional information in the lower-dimensional representation; in graph layout this is often the graph-theoretic distance between pairs of nodes.  
Multidimensional scaling (MDS)  is commonly applied to layout and involves minimising the difference between the Euclidean and graph-theoretic distances. There are two MDS approaches that solve this problem: the more common distance scaling and classical scaling.

Distance scaling aims to minimise the sum of squares of the difference  (the dissimilarity) between the graph-theoretic and the Euclidean distance for each pair of nodes in the layout, known as the stress, through an optimisation procedure. Higher stress indicates a poor representation of the original distances between nodes in the layout~\cite{Freeman_2005}. Minimisation of the stress is achieved through a statistical technique of stress majorisation~\cite{Gansner_2004}. Distance scaling has been used for social network layout~\cite{Kruskal_1980}, co-worker relationships~\cite{Freeman_2005}, and as part of the XGvis system~\cite{Buja_2008}. 

Two recent dimension reduction based techniques that incorporate multivariate data are EdgeMaps by Dörk \emph{et al.}~\cite{Dork_2011} and the projection explorer graph, PEx-Graph~\cite{Martins_2012}. The aim of EdgeMaps was to unite the explicit and implicit relationships though an MDS based layout that computes attribute similarity. Position is double encoded by hue and saturation. Only one node's links are shown at a time for readability but this restricts exploration to one node's relationships,  thus the similarity between two nodes cannot be easily explored. Further, the layout is not interpretable in terms of its attributes. 

PEx-Graph extends the Projection Explorer tool~\cite{Paulovich_2007} for exploring high-dimensional data projections. PEx-Graph aims to visualise heterogeneous graphs using projections based on either attribute data, connectivity data or both, working from the principle that node proximity should equate with node similarity. Dimension reduction is through the IDMAP technique~\cite{minghim_2006} and one of the aims is to improve clustering and reduce visual clutter. 

In this paper we utilise the dimension reduction tool TPP (targeted project pursuit) for graph layout. The next section introduces the algorithm used in TPP and discusses how this has been adapted for graph layout in an updated tool known as graphTPP. 

\section{Targeted Projection Pursuit and graphTPP}\label{sec:tppintro}

Targeted projection pursuit (TPP)~\cite{Faith_2006, Faith_2007} is a linear dimension reduction method for exploring high-dimensional data spaces based on projection pursuit (PP). PP aims to find the most interesting projection. `Interestingness' is based on a projection index that optimises a particular feature of the data. One such feature is class separation~\cite{Lee_2005}.  TPP takes this idea one step further by allowing the user to define their own notion of what is interesting and then seeks to find a matching projection. The technique is implemented in an interactive tool,  with a drag and drop interface, that allows the user to select groups of points and move them around the 2D projection space. The user is then presented with the closest possible projection to the one they desire.  This projection can then be used to identify outliers, possible misclassifications and feature contributions  as well as visualising the classification~\cite{Faith_2007}.

\subsection{Targeted Projection Pursuit}\label{sec:tpp}
In TPP the model takes an $n \times p$ matrix $X$ with $n$ nodes and $p$ attributes (just the numerical attributes, not those that define cluster membership). When a user drags a point, or set of points, they are defining a target view $T$ that represents their expectation about how the data should appear. The target view is an $n \times 2$ matrix which describes the target positions of the nodes in 2D. The aim of TPP is to find a $p \times 2$ projection matrix $P$ that minimises the difference between the target view $T$ and the projection of the original data $XP$ as in Equation (\ref{eq:tpp}).

\begin{equation}
\text{min} \left\| T- XP \right\| 
\label{eq:tpp}
\end{equation}

Equation (\ref{eq:tpp}) is solved by training a single layered perceptron with $p$ inputs (the attributes) and two outputs (the target view) through the standard back-propagation algorithm of an artificial neural network. Attribute data for each of the $n$ nodes is presented in turn and the network is trained to produce the corresponding row of $T$ according to a least-squares calculation. In this case the entire training set is the testing set and once convergence is reached the original data can be transformed into the 2D view where the connection weight between each input neuron and the two output neurons (for the 2D view) gives the weight of each attribute in the projection. The TPP tool itself is built upon the data mining software Weka~\cite{hall_2009} and is written in Java.\footnote{\url{https://code.google.com/p/targeted-projection-pursuit/}}

There are two ways users can interact with the data: an automated separation process or direct user interaction. The goal of automated separation is to try to move each cluster as far from each of the other clusters as possible. This is done by defining the target projection to be the projection of the $k$-simplex onto 2D space, associating each point of the simplex with a cluster, and then moving each cluster towards the points of the simplex. Manual separation follows the same principle but instead the user defines the target projection by interacting with the points directly. Exploring a dataset usually requires a combination of both actions. Hu et al.\cite{hu_2013} also considered interactions similar to those in TPP but instead applied their methods to an MDS algorithm that translates the user's actions into adjustments of the parameters of the algorithm. 

\subsection{graphTPP}\label{sec:gtpp}

For the purposes of this research the TPP tool was extended to incorporate graph layout, and is henceforth referred to as graphTPP. This section describes the features added to graphTPP to support graph layout. 

The key new feature for graphTPP is the ability to import and then display the edges that define the relationships of the graph. This can either be achieved through importing a separate `edges' CSV file which defines the source node, target node, and, if required, an edge weighting. Alternatively, the edges are imported as part of Weka's native ARFF file whereby the edges are defined as an adjacency matrix. 

Once imported, edges can be displayed as straight lines (with optional arrowhead indicating direction) or curved lines where following the edge clockwise indicates the direction. A further option for displaying edges as bundles is introduced in the \nameref{sec:bib} section. Edges can be either be coloured neutrally (grey)  or based on the colours of the nodes they connect. graphTPP also supports filtering of edges to improve clarity. When a group of nodes is selected the edges connected to these nodes maintain their opaque appearance and other edges become more transparent, the level of which is controlled by the user. A user can even elect to hide the edges not connected to a currently selected node. Edges can also be filtered so that only incoming/outgoing edges are displayed and, if an edge weight has been defined, a user may also filter the display of edges based on these edge weights. 

\begin{figure*}[htbp]

	\centering
	\includegraphics{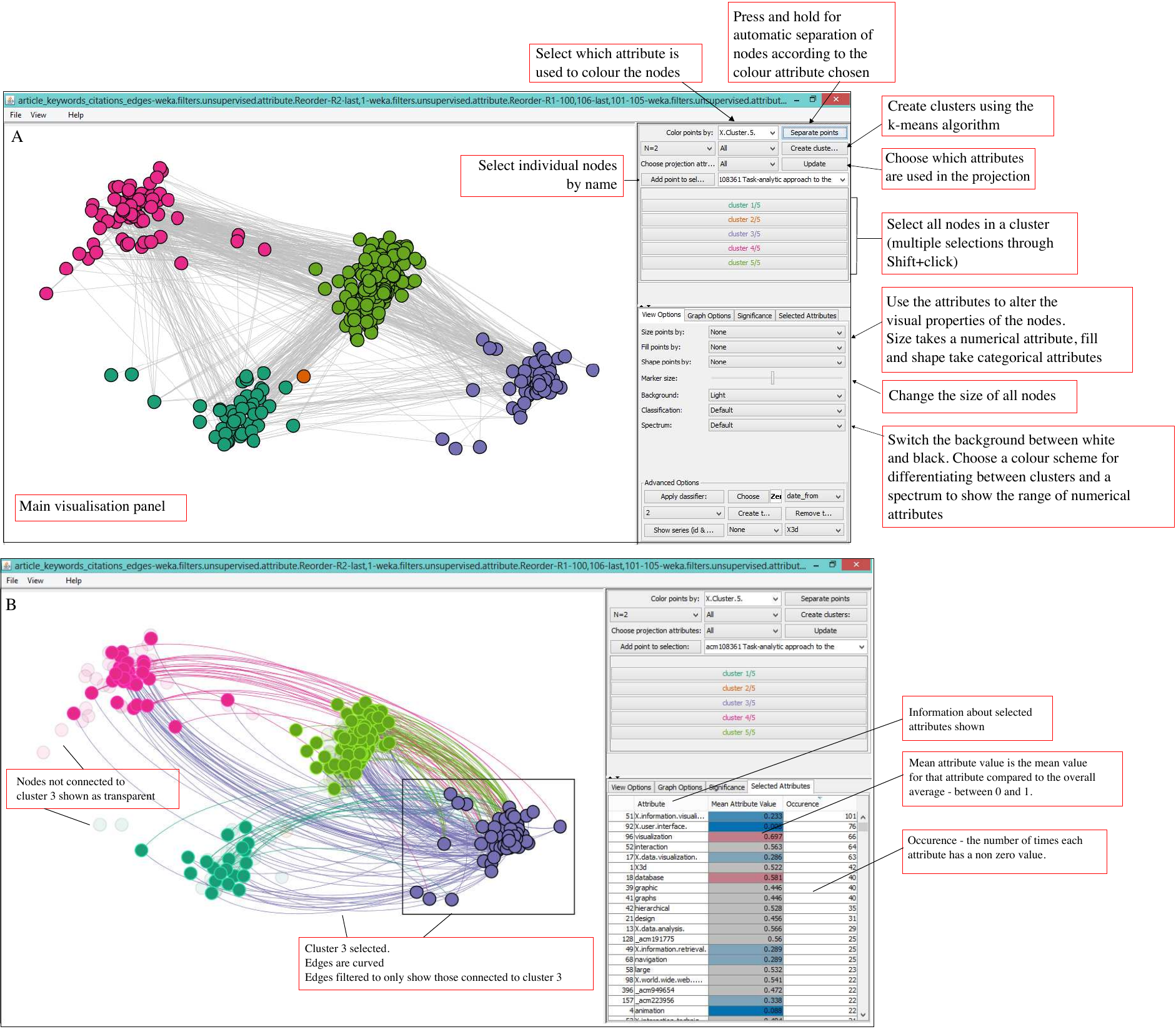}
	\caption{{\bf The graphTPP interface with a node-attribute graph already imported} (A) shows the panel which can be used to perform automated separation as well as some overall colour options. (B) shows a graph where curved edges have been selected and coloured by their source node. A specific cluster has been selected which makes nodes not connected to that cluster appear as transparent.}
	\label{fig1}

\end{figure*}

In the main visualisation panel (see Fig.\ref{fig1}) a pan and zoom interaction is provided that enables the user to move around the projection space and zoom into clusters of nodes.  The colour of a node can be based on cluster membership (using one of  ColorBrewer's~\cite{colorbrewer} six colour schemes), the amount of node transparency depends on whether the node is selected or not (e.g., Fig.\ref{fig1}(B)), nodes can be sized according to an attribute's value or by its in/out/overall degree, and a node's label may also be displayed and can be filtered or sized based on node degree (not shown). 

The significance panel (see Fig.\ref{fig1}(B)) is similar to the original TPP version. Each column shows the weight applied to each attribute in both the $x$ and $y$ directions corresponding to the current projection. If the axes are displayed on the main visualisation panel then selecting a row in this table will highlight its corresponding axes while right-clicking a row will colour all the nodes according to the value of that attribute.  If, during the analysis, the user suspects that some nodes are irrelevant or noise they can elect to remove them from the projection calculation and the layout will recompute with only the remaining attributes. 

An additional panel shows an extension of the axes colouring that occurs when nodes are selected. If the axes are displayed and a group of nodes is selected, the colour of each of the axes will reflect the comparison between the overall mean value of each attribute to the mean value of the attribute restricted to the selected nodes. The selected attributes table provides a representation of that data in tabular form which also displays the number of times each attribute has a non-zero value. This table can also be sorted and nodes can be coloured directly by again right-clicking on the desired attribute.

graphTPP and all data used in this paper can be downloaded from \url{https://github.com/helengibson/graphTPP/}.

\section{Case Studies}\label{sec:cs}
In this section we present three case studies that showcase the different features of graphTPP. 

\subsection{Influence Graphs}\label{sec:inf}
Influence graphs have directed edges indicating the influence one node has had on another. In this case the nodes are artists and an edge represents one artist's influence on another. The purpose of visualising an influence graph is to explore the impact of these influences and how they correspond to the given attributes. 

The dataset was originally sourced from Freebase (\url{freebase.com}) for the EdgeMaps application~\cite{Dork_2011}\footnote{\url{http://mariandoerk.de/edgemaps/demo/}} and is reused here to provide a comparison between the two applications. Edgemaps allows users to explore the relationship between explicit and implicit information spaces. The dataset contains influence relationships between 226 artists  with 281 directed links in the direction of artist A influenced artist B. There are 168 binary attributes describing features such as profession, nationality, art forms, art periods and movements.

In this section we show (1) the use of a principal component analysis (PCA) projection as a starting point for data exploration, (2) how generating clusters through $k$-means clustering  focuses the exploration of previously unclustered data, (3)  how direct interaction with the graph leads to new insights,  (4) how combining attributes with the graph enables a richer visual analysis of the structure of the graph, and (5) how is it not possible to make these same  insights with other layout techniques.

\subsubsection{PCA Layout}

\begin{figure*}[htbp]
	\centering
	\includegraphics{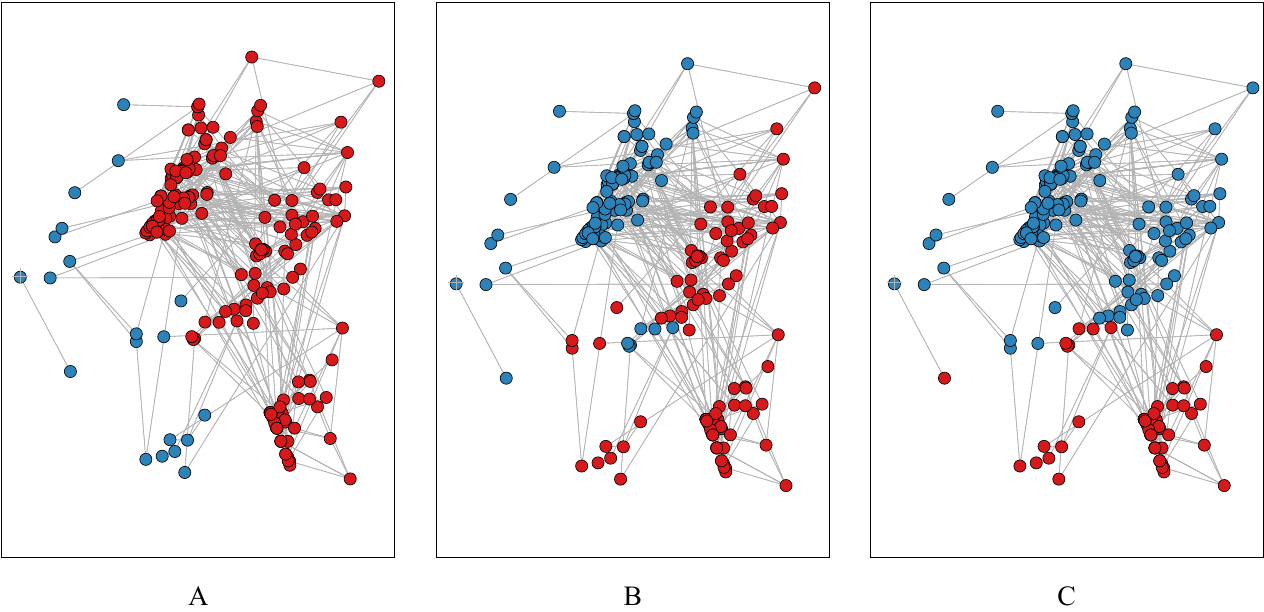}
	\caption{{\bf The PCA view of the artists' influences dataset coloured by the presence of three different attributes with red indicating presence and blue absence of the attribute for that particular node.} (A) Painting, (B) Artist, (C) United States }
	\label{fig2}
\end{figure*}

Initially graphTPP presents a PCA projection layout that provides a starting point for investigating the graph's structure (Fig. \ref{fig2}). The layout shows four potential clusters, plus a number of peripheral nodes on the left-hand side. 

Fig. \ref{fig2} shows the graph coloured by the three most significant attributes in the PCA projection: Painting (Fig. \ref{fig2}(A)), Artist  (Fig. \ref{fig2}(B)) and the United States  (Fig. \ref{fig2}(C)).  These three attributes define the four main clusters. The bottom right cluster has all three attributes, the bottom left cluster has the attributes United States and Artist, the middle cluster has Painting and Artist and the uppermost group has only Painting. This demonstrates that (1)  those nodes with the United States attribute are split into two groups based on whether they have the Painting attribute, but they do not influence one another; (2) those with none of the three attributes are only sparsely connected to the rest of the graph and two nodes not at all. This indicates a relationship between the attributes, the graph and the layout.

\subsubsection{Clustered Layout}
A given clustering does not always exist but graphTPP's  strength is in separating clusters visually in order to spot patterns. Thus without a clustering users can feel lost in terms of where to begin their interactive exploration. This problem can be alleviated by the computation of a clustering. In graphTPP, $k$-means clustering, based on the Euclidean distance measure, is implemented. In $k$-means clustering it is up to the user to define the number of clusters to be created, which is not always clear. However, the process of determining this can encourage exploration of the graph. This ultimately improves the user's understanding of the dataset, as they construct cluster-based `what-if' scenarios to elicit insight from the graph. Next we present a three cluster layout of the influence graph while demonstrating the automatic and manual interaction techniques available in graphTPP. 

\begin{figure*}[htbp]

	\centering
	\includegraphics[width=\textwidth]{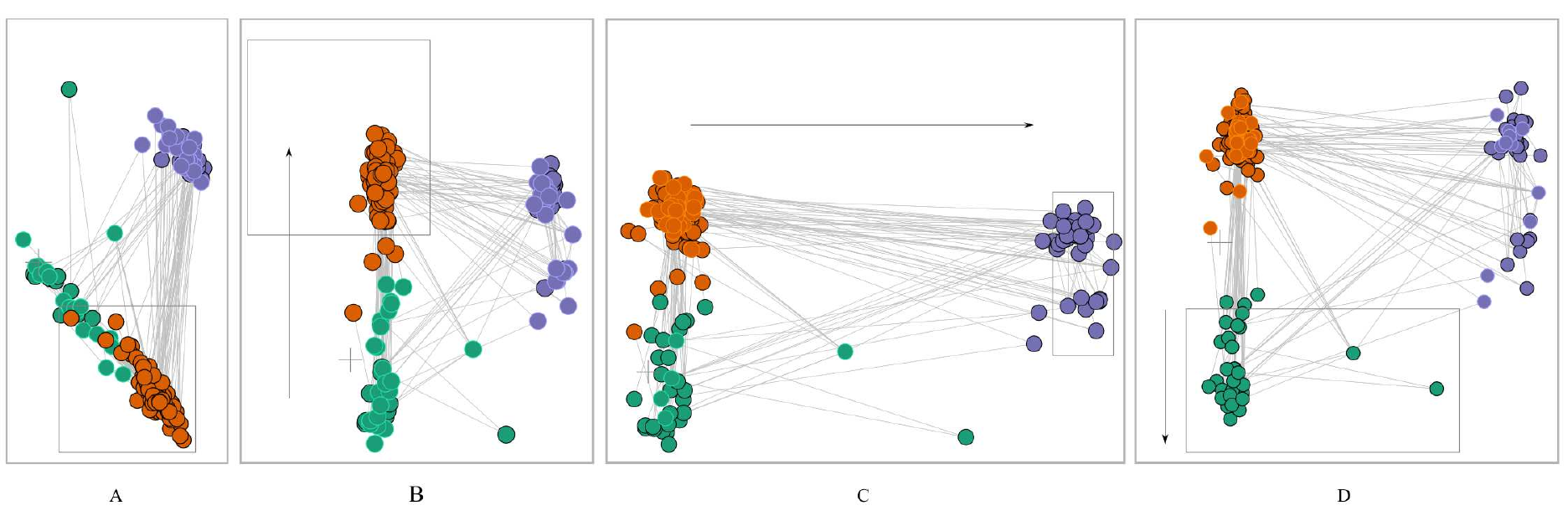}
	\caption{{\bf The interactive discovery process with graphTPP.} The three clusters are initially automatically separated (A). The user attempts to improve the visual separation between the two joined clusters by (B) dragging the orange cluster to the top causing the purple cluster to divide, (C) dragging the purple cluster to the right, and (D) pulling the green cluster down to accentuatethe separation in the purple cluster.}
	\label{fig3}

\end{figure*}

Fig. \ref{fig3}(A) shows the layout of the graph after automatic separation of the clusters. One group (the purple group) is more separable than the other two which overlap. The natural exploratory step is to separate these two overlapping clusters; firstly by dragging the orange cluster up (Fig. \ref{fig3}(B)), then increasing the separation distance of the purple group  (Fig. \ref{fig3}(C)) and finally by pulling the green cluster downwards (Fig. \ref{fig3}(D)). Through these steps three things become apparent: (1) a sub-division is forming in the purple cluster; (2) the nodes in the green cluster are mostly connected to the nodes in the upper half of the purple cluster; and (3) two green nodes appear as outliers.

\begin{figure*}[htbp]

	\centering
	\includegraphics[width=\textwidth]{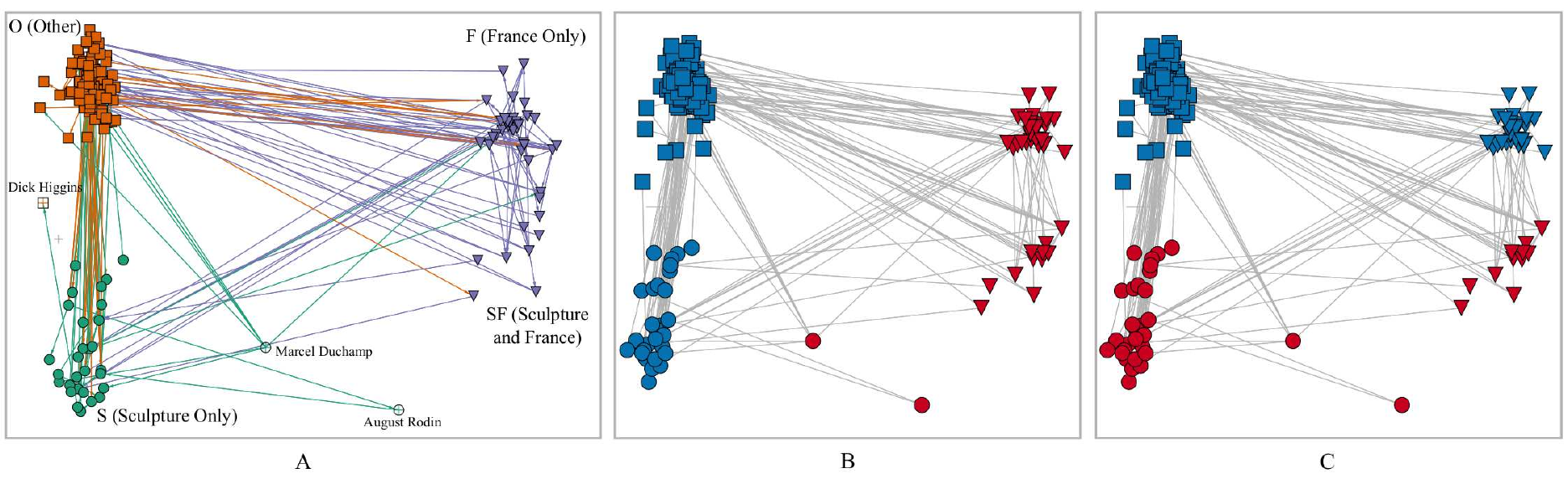}
	\caption{{\bf The influence graph layout with three clusters and attribute distribution.} In fact, four clusters emerge with the lower two containing nodes with the attribute Sculpture and the right-hand side those with the attribute France (A). The three named nodes are potential outliers. (B) and (C) show the distribution of the France and Sculpture attributes in the three cluster graph.}
	\label{fig4}

\end{figure*}

In these three clusters,  two attributes, Sculpture and France, are found to be the most influential and the four combinations of these attributes represent the clusters in the layout. These groups are abbreviated to O (other -- has neither the Sculpture or France attribute -- orange square),  S (Sculpture attribute  only -- green circle), F (France attribute only) and SF (Sculpture and France attributes) both blue triangles as shown in Fig. \ref{fig4}(A). Three nodes, Dick Higgins, Marcel Duchamp and Auguste Rodin appear as outliers.  Figs. \ref{fig4}(B) and \ref{fig4}(C) show the distribution of the two attributes, France and Sculpture, over the whole graph.

This layout gives us some insights into the relationships between the attributes and the graph's structure. 

\begin{itemize}
	\item Between the Other and the Sculpture clusters the numbers of incoming and outgoing edges are similar, while to the France and SF clusters there are more incoming than outgoing edges. This implies those with the attribute France are mostly influenced by others who share that attribute. 

	\item There are only three influence relationships from Sculpture cluster to the whole France cluster: two to the France only cluster and one to SF cluster. 

	\item Those in the SF cluster are not influential within their own cluster and are influenced by others with the attribute France rather than Sculpture. 

\end{itemize}

The three possible outliers detected in this layout are Auguste Rodin (has both France and Sculpture attributes but placed in the Sculpture only group due to his profession attribute of Sculptor), Marcel Duchamp (has the France attribute but also the United States attribute which pulls his node left) and Dick Higgins (has only one influence with which he shares two attributes Artist and United States). 

This demonstrates how the interaction step is  as important as  automatically separating the points in producing the most informative layout. It was the investigation and the search for a clearer layout (i.e., could the orange and green clusters be separated?) that alerts the user to the possibility of two sub-clusters within the purple cluster. The user can then investigate the attribute composition of these groups and the edges of the graph itself. Exploration can continue by increasing the number of clusters generated and repeating this process. 

\subsubsection{Comparison to other layouts}

\begin{figure*}[htbp]
	\centering
	\includegraphics{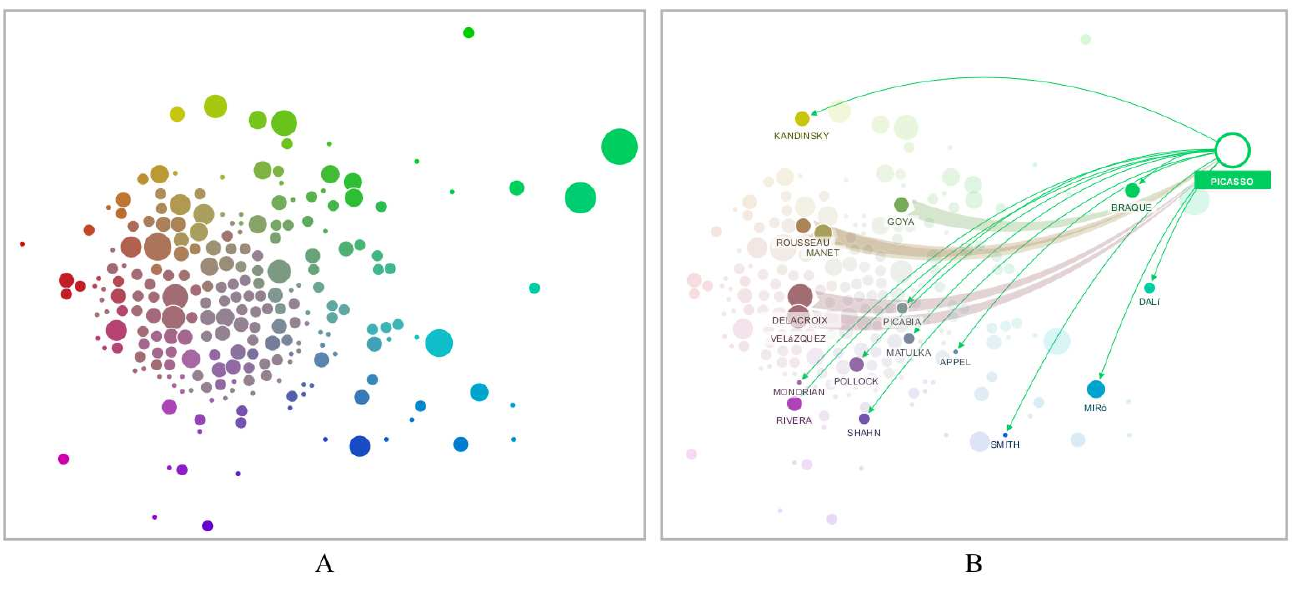}
	\caption{{\bf Layout of the artists influence graph using Dörk \emph{et al.}'s EdgeMaps~\cite{Dork_2011}}  (A) shows the initial layout and (B) the selection of a node.}
	\label{fig5}
\end{figure*}

EdgeMaps~\cite{Dork_2011} is a web-based network visualisation that visualises explicit and implicit relationships in networks to integrate multiple relationships, show invisible data patterns, and support serendipitous data explorations. Explicit relations are edges and implicit relations indicate the attribute similarity based on MDS.  The MDS result provides a  meaningful visual variable where similar nodes are placed close together and dissimilar nodes further apart.  EdgeMaps redundantly encodes influence through edge size and visibility; meanwhile, attribute similarity is encoded through colour and position, as in  Fig. \ref{fig5}(A).

Interaction is via a pivot-based technique where the user clicks from node to node following the influence relationships; however, as can be seen in Fig.~\ref{fig5}(B) only one node's edges are displayed at any one time meaning a user can only make comparisons between nodes from memory. Only minimal exploration of why nodes may be placed close together is possible thus some of the advantages gained by integrating these implicit and explicit relationships are lost.

\begin{figure*}[tbp]

	\centering
	\includegraphics[width=\textwidth]{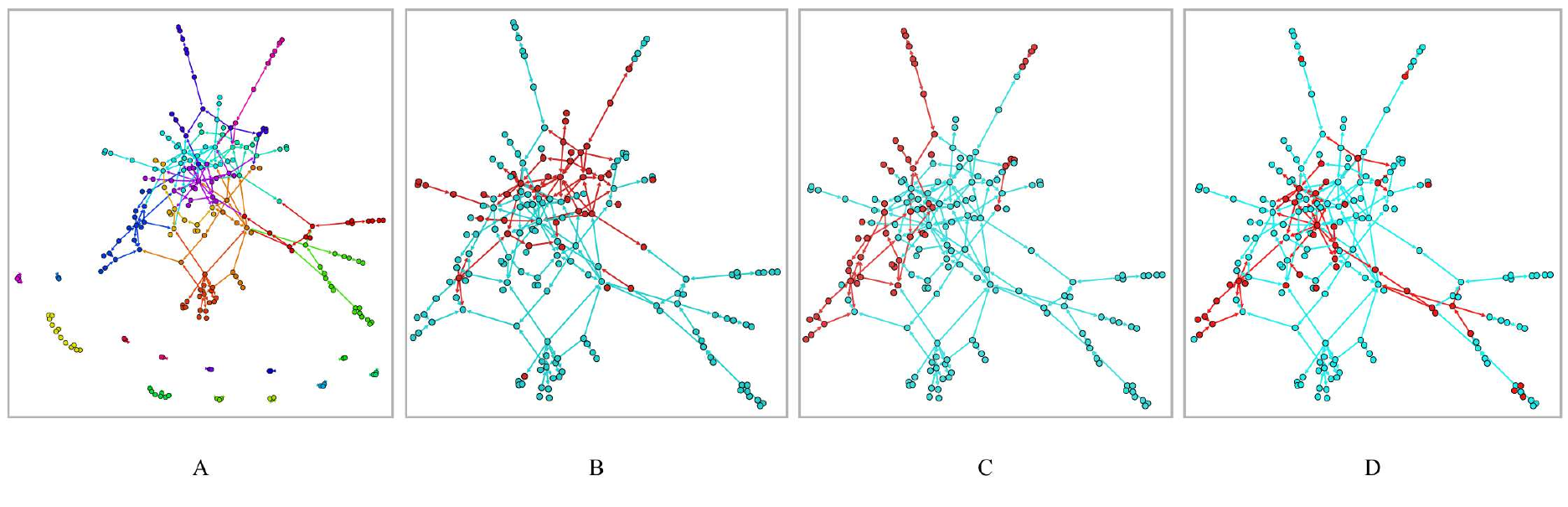}
	\caption{{\bf The layout of artist's influence graph using the Force-Atlas layout in Gephi.} Nodes in (A) are coloured according to a community detection algorithm and in (B) to (D)  by attribute (red) for the France, United States and Sculpture attributes respectively.}
	\label{fig6}

\end{figure*}

Force-directed techniques use the connectivity structure of the graph  for layout placing connected nodes close to one another. Their layouts are often criticised for not providing meaningful insights to the user.  For this dataset there is one immediate insight from a force-directed layout that is not apparent with graphTPP or EdgeMaps: the graph has one giant component and many small disconnected groups (Fig. \ref{fig6}). 

However, beyond these disconnected components it is difficult to learn more from this layout. Figs~\ref{fig6}(B) to \ref{fig6}(D) show the distribution of some significant attributes from the graphTPP layout but even the relationships between these attributes and the edges are difficult to detect. 

Overall, graphTPP and EdgeMaps both make good use of attributes for layout while the major advantage of the force-directed layout is the separation of the disconnected components. In EdgeMaps the nodes are spread to remove overlap, in the force-directed layout most nodes are visible although there is some bunching in the centre, while in graphTPP the clustering causes significant overlap. This indicates that the clustering is strong but obscures within-cluster edges; however, having nodes distant from their cluster highlights an outlier. Both EdgeMaps and the force-directed layout make it difficult to interrogate the underlying relationships between the attributes while this is a key feature of graphTPP allied with its interactivity and clustering. The interactivity allows the user to find interesting projections themselves and explore and investigate their hypotheses. Having found a layout, be it a clustered one or otherwise, graphTPP then shows a strong structural overview of the graph according to that projection assisting the user in investigating significant attributes and in constructing a model of how the graph's structure and the attributes are related. 

The next section will extend graphTPP to a larger dataset based around a real-world problem of network security.

\subsection{Network Security}\label{sec:netsec}

The VAST challenge runs each year as part of the IEEE Conference on Visual Analytics Science and Technology (VAST). Participants are provided with data and challenged to develop a visual analytics tool in order to discover a ground truth present in the data. The data are artificial but represent a real-world scenario. In 2012 the scenario was a network in a regional office of the fictional Bank of Money (BoM).%
\footnote{This dataset was taken from the VAST 2012 challenge at \url{http://www.vacommunity.org/VAST+Challenge+2012}.}
The office has recently been upgraded to support a 24 hour call-centre and is experiencing difficulties with its network and computers with staff reporting pop-up messages about spy-ware, viruses and illegitimate anti-virus software; constantly running hard drives; and slow performance. Intrusion detection system (IDS) and firewall logs were provided for two days worth of data. The aim is to identify security concerns, trends in the logs and the root cause of the problems. The following sections report on the steps taken to discover the events of concern through graphTPP. 

\subsubsection{The IDS logs}\label{sec:ids}

An IDS monitors network activity according to a set of predefined rules. When a rule is broken an alert is fired with the relevant information. The IDS used by the BoM network is snort~\cite{roesch_1999} configured with the Emerging Threats ruleset.\footnote{\url{http://rules.emergingthreats.net/open/snort-2.9.0/}} The data for the IDS logs is split into two days covering the period from 2012/04/05 17:55 until 2012/04/06 17:23 and from  2012/04/06 17:23 until 2012/04/07 08:59. For each alert a number of data points are recorded including source and destination IP address (the nodes) and port number, classification (one of five), priority (three levels), a label (one of 21), packet info and xref (usually some URLs corresponding to the type of threat).  Node attributes are defined by counting the number of each edge type connected to each node. Initially direction was not considered but the second analysis doubled the number of attributes in order to include direction. Each node was classified according to its IP address and the attributes were  standardised to the interval $[0, 1]$ so that less frequent, but important, events still show up.

\subsubsection{First day, combined source and target events}

\begin{figure*}[tbp]
	\centering
	\includegraphics[width=0.75\textwidth]{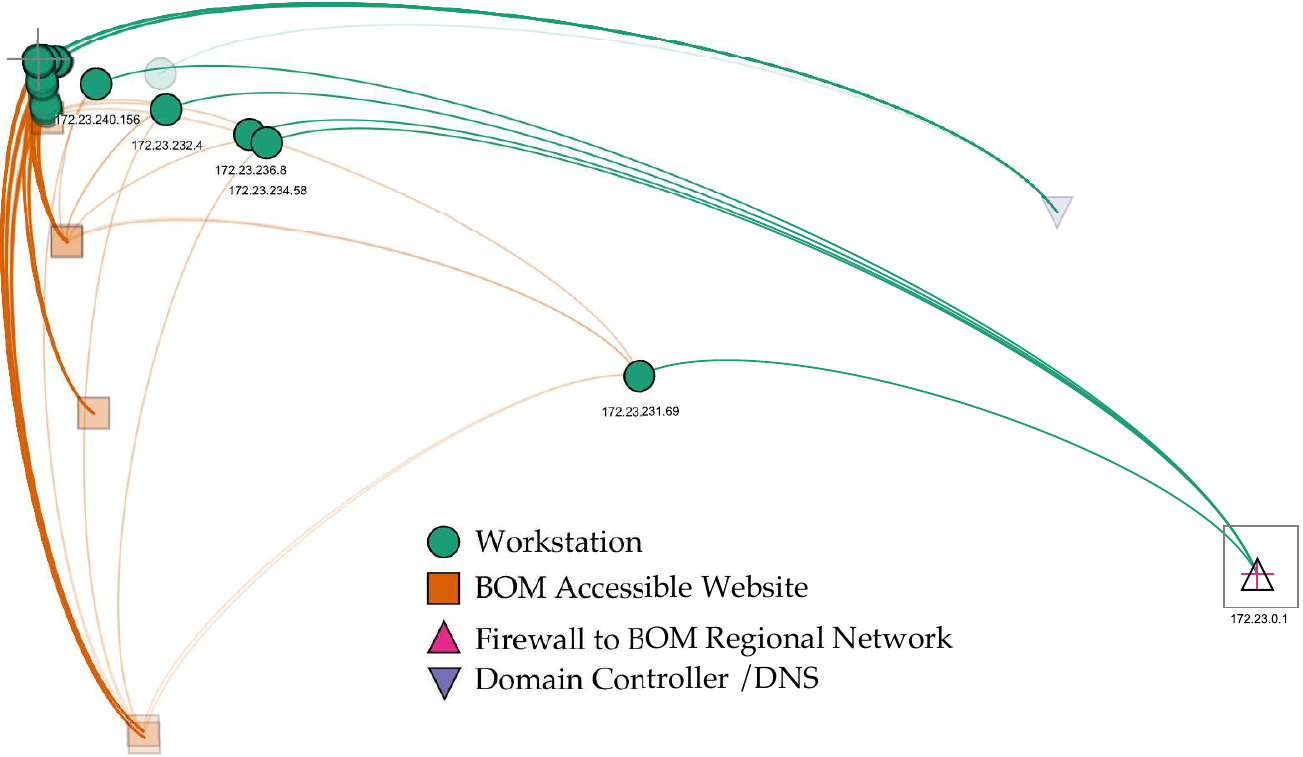}
	\caption{\bf A filtered graphTPP layout of the IDS logs graph for the first day which  highlights the five  green labelled  workstations that connect to the firewall to the regional BoM network (the selected pink triangle).}
	\label{fig7}
\end{figure*}

On the first day four node classifications are detected: workstation, domain controller/DNS, BoM accessible website and the firewall interface to the regional bank network. After automatic and manual separation most BoM accessible websites and workstation nodes are clustered in the top, left-hand corner. Fig. \ref{fig7} shows six workstation nodes (five are highlighted and one is shown with some transparency) that appear further to the right than the other workstation nodes along with a DNS and firewall node. This indicates that the attributes of those six nodes differ compared to other workstations, further evidenced by the structure of the graph itself. The most significant attributes in this projection are `Misc Activity' and `ET POLICY IRC authorization'. These are common events and appear to be when a BoM accessible website connects to a workstation (as will be seen in the solution this is actually representing the spread of the infection).

Five of the six workstation nodes are labelled with their IP address and their edges are fully opaque. These five nodes are 172.23.231.69, 172.23.234.58, 172.23.236.8, 172.23.232.4 and 172.23.240.156. They were also identified as outliers by Harrison \emph{et al.}~\cite{Harrison_2012} whose streaming graph system \textsc{situ} scores events and presents them as cards to the user. Using the attribute distribution table, graphTPP can then identify that only these five nodes and their target node, the firewall interface to the BoM regional network, have attributes which are involved in events accessing the database ports, VNC scans, and  SNMP (simple network management protocol) requests over TCP. 

For this challenge, graphTPP was linked to a MySQL database so that selecting a node allowed the user to bring up the raw log entries mentioning that node. This allows graphTPP to act, in part, as a visual search interface to the database containing the raw data which may be useful to a network analyst who would then be able to focus their efforts on these sections of the logs.  

Searching for the outlier workstations shows the `xrefs' in the log contain URLs  linking to  information about the detected threats. Several URLs are associated with the SNMP alert indicating that a denial of service attack or remote execution of code may be initiated on a  user's machine.\footnote{\url{http://cve.mitre.org/cgi-bin/cvename.cgi?name=2002-0013}}\footnote{\url{http://cve.mitre.org/cgi-bin/cvename.cgi?name=2002-0012}}\footnote{\url{http://www.securityfocus.com/bid/4132}}\footnote{\url{http://www.securityfocus.com/bid/4089}}\footnote{\url{http://www.securityfocus.com/bid/4088}} Variations on this pattern involve an SSH scan which may indicate a brute force attack. (As seen in the solution in the next section, these SSH scans are attempts to exfiltrate data from the system.)

The node with IP address 172.23.231.69 is even further to the right indicating it may have additional attributes which are found to be SSH scans: events which target the email ports and raises an event related to unusually fast terminal server traffic indicating the presence of a worm.\footnote{\url{http://threatpost.com/en_us/blogs/new-worm-morto-using-rdp-infect-windows-pcs-082811}} This would attack through a remote desktop protocol on TCP port 3389, an event which is also present.

A sixth outlier (IP 172.23.5.110), which appears just above node 172.23.232.4 in Fig. \ref{fig7}, connects to the domain controller/DNS. Whilst this is not the only node that connects to the DNS, it is the node which does it significantly more often than other nodes and is related to the NETBIOS.

Analysing the logs for the second day yields little further information although there are no connections to the firewall on this day. 

\subsubsection{Separated source and target events} 
\begin{figure*}[htbp]
    \centering
    \includegraphics{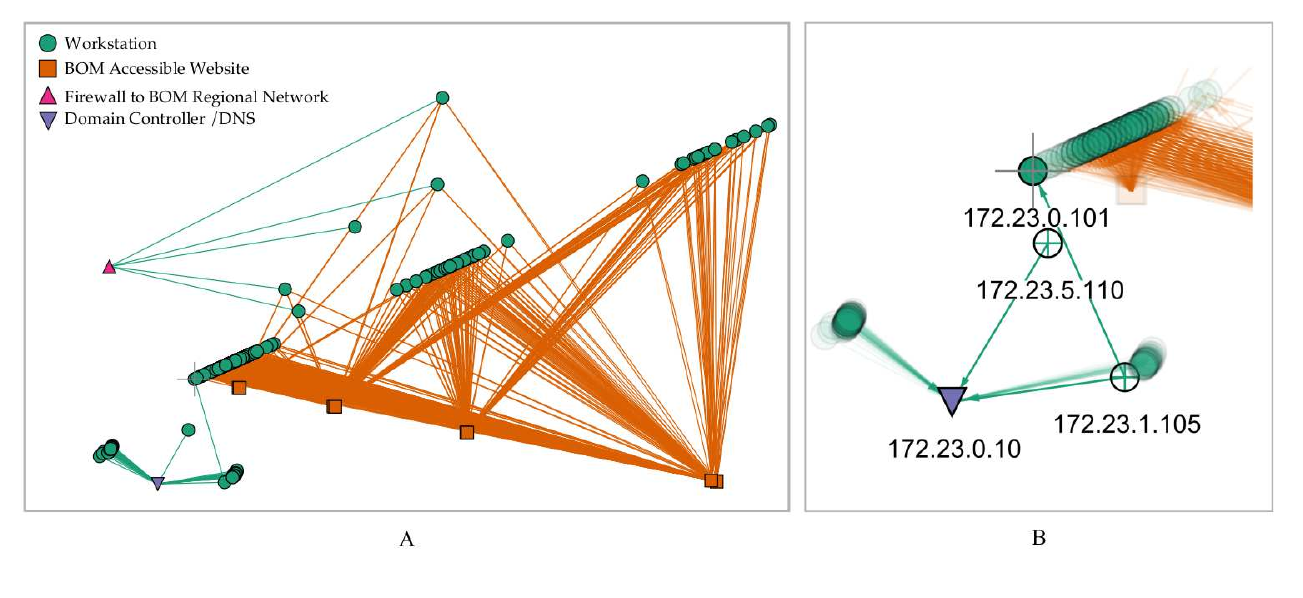}
    \caption{{\bf graphTPP layout of the IDS logs for the first day with separated source and target attributes.} Both automated and manual separation of the nodes was used in creating the layout. (A) displays the whole graph layout for the IDS logs while (B) provides a close up of the nodes connecting to the DNS.}
    \label{fig8}
\end{figure*}

When we consider whether a node is acting as a source or target node we double the number of attributes under consideration. Now, for the first day the layout in Fig. \ref{fig8}(A) can be created. This is is similar to our original day one graph with the main five outlier workstation nodes appearing in the upper part of the graph. There are also two further clusters of workstation nodes that connect to the domain controller/DNS and three groups that appear along one axis indicating a differing number of connections made in that day.

For the two groups that connect to the domain controller/DNS the group to the left hand side is involved  in NETBIOS events, while the group to the right is involved with a Potential Corporate Privacy Violation and an ET POLICY DNS Update From External net. In this layout two nodes stand out. One is a node (with IP address 172.23.5.110) which belongs to neither group but still connects to the domain controller/DNS. The second is a node  (IP address 172.23.1.105) with one outgoing link that connects to a second workstation (with IP address 172.23.0.101) (Fig. \ref{fig8}(B)). This is  the only occurrence in the IDS log where two workstations connect to one another; however, it is difficult to ascertain the significance of this event. 

Again, the graph for the second day was less informative and showed two components: one where workstation nodes connected to BoM accessible websites and a second where workstation nodes only connected to the DNS. 

\subsubsection{Firewall logs}\label{sec:firewall_logs}

The firewall logs cover the same timescale as the IDS logs. For each  connection through the firewall a number of attributes are recorded including the source and destination IP address and port number as well as a timestamp, syslog priority, operation and message code. The dataset was initially parsed in a similar fashion to the IDS logs: compiling the total number  of occurrences for each possible event on each node and then standardising the values for each attribute. However, it was not possible to produce a layout with graphTPP due to the size of the data (approximately 3500 nodes, 60 attributes, 50,000 edges) which restricted interactivity and resulted in an unreadable projection.

An alternative method for constructing attributes for the firewall data is to use the number of connections per fifteen-minute interval to highlight nodes with similar patterns of activity. Nodes were classified according to their IP address which identified an unclassified group with IP addresses in the range 172.28.X.X which is not listed as part of the BoM network. 

\begin{figure*}[htb]
    \centering
    \includegraphics[width=0.7\textwidth]{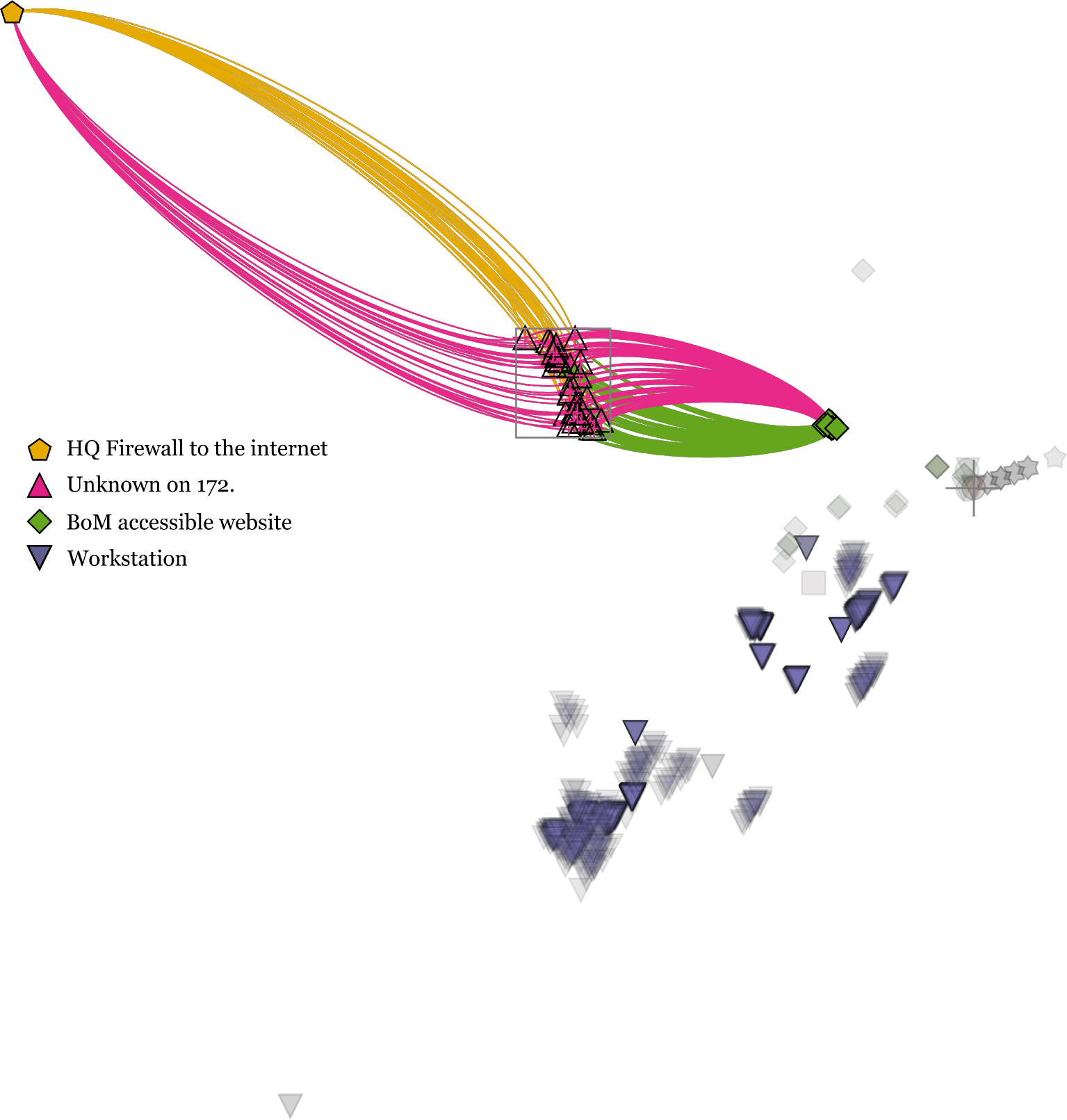}
    \caption{\bf Layout of the firewall graph minus the outliers only showing the connections of those nodes that are unknown in IP range 172.28.X.X.}
    \label{fig9}
\end{figure*}

An initial graphTPP layout highlighted two outlier nodes with IP addresses 172.23.252.10 and 17.23.0.132 that have much higher levels of activity than any other nodes in the network: one node has high activity between the times 06/04/2012 17:55 and 07/04/2012 09:10 and while the second outlier spikes in use between 06/04/2012 15:10 and 06/04/2012 17:10. Further, for this latter node many of the connections are to port 6667. This port is commonly used for IRC (internet relay chat) which can be exploited by hackers. In the firewall logs there are more than 30,000 instances of port 6667 being used as a source port and over two million as a destination port. However, the extremeness of the positioning of these two nodes in the layout made analysis of the rest of the graph infeasible. Fig. \ref{fig9} shows the graph with these two nodes removed and with some edges filtered out concentrating on the nodes the unknown workstations connect to (the HQ firewall and a specific group of BoM accessible websites). This group both sends and receives connections while other workstation nodes do not, although these returning connections were actually denied. Nevertheless, there should not be unknown machines connected to the network. These connections appear for a short period of time: from 05/04/2012 18:22 to 06/04/2012 00:27  and  from 06/04/2012 18:06 to 07/04/2012 00:57, a similar time period each night.  

The firewall logs were less informative than the IDS logs and the layout was more difficult to interpret. The volume of nodes also contributed  to slow loading times and a delayed interaction meaning the interactive features of graphTPP were diminished. The analysis did, however, lead to the discovery of some unknown nodes on the network and their unusual interaction pattern. 

\subsubsection{Comparison with the ground truth}\label{sec:security_groundtruth}
At the end of the contest the ground truth was published. The goal was to identify the introduction of a botnet which spread and exfiltrated data from the network to external machines. Initially ten command and control servers come online (the BoM accessible websites). A client connection from an infected workstation to these servers activates the botnet. The server scans the workstation to determine its type (through checking email and database ports). If it finds a server the botnet looks for sensitive data to exfiltrate through SSH or FTP protocols; otherwise, if it is a workstation, the botnet port scans other network computers for vulnerabilities. The botnet clients also demonstrate a set of behaviours complementary to those of the command and control servers.

In comparison to graphTPP the ground truth presents an overall scenario whilst graphTPP highlighted key nodes and events.  For example, graphTPP identifies communication between the workstations and the command and control servers but the significance of this was not initially identified. graphTPP also showed which workstations had been targeted in an attempt to access data from the email or database ports and the use of SSH. graphTPP did not support the analysis of the firewall data well. In particular, variations in the number of connections as the IT department attempted to combat the attack were not detected. The firewall attribute view also does not highlight the attempts to access the data through FTP nor that those attempts had been denied. 

graphTPP's strength was in facilitating the identification of individual nodes acting unusually. The main difficulty was the inability to incorporate temporality in the display making it hard to identify overall trends. This not only applies to the edges, and the time of a connection, but also because only the cumulative effect of the attributes is ever seen thus it is difficult to distinguish between what is normal and what is unusual. That is, while individual events may be pinpointed, overall trends over the days may not be. Therefore, while it has been shown that graphTPP can be used as a graph layout method for visualising network security logs, it may be more useful as part of a wider system than as a stand-alone tool.

\subsection{Bibliographic Data}\label{sec:bib}

Citation and bibliographic networks are  common in network visualisation as they can facilitate a rapid understanding of a research area, discovery of key papers, authors and publication venues, identification of trends, emerging areas, and the formations of communities within the research landscape~\cite{dunne_2012}. Software tools that support visualisation of bibliographic networks include HistCite~\cite{Garfield_2004}, Eigenfactor (\url{eignefactor.org}), CiteWiz~\cite{Elmqvist_2007}, and  Action Science Explorer~\cite{dunne_2012}. PivotPaths~\cite{dork_2012} allows exploration of a subset of the Microsoft academic search database and many generic network visualisation tools support the visualisation of bibliographic networks with the main challenge being the collection, formatting and input of data. 

\subsubsection{The History of InfoVis}

The \emph{History of InfoVis} contest challenged participants to track the development of information visualisation through using citations and metadata such as keywords, authors, abstracts and year of publication for articles published at the InfoVis conferences.\footnote{\url{http://www.cs.umd.edu/hcil/iv04contest/info.html}} Additionally, cited papers that form part of the ACM digital library were  included but with incomplete metadata.

The dataset used was a processed version from Indiana University which was further cleaned to consolidate the number of keywords to 1390.\footnote{\url{http://iv.slis.indiana.edu/ref/iv04contest/iv04-contest.mdb}} A paper represents a node, citations are edges, and keywords are attributes; with the condition that a keyword must be an attribute in at least six different articles, those articles without keywords were removed.  This resulted in a graph with 100 attributes describing 395 nodes and 989 edges. 

\begin{figure*}[htbp]
	\centering
	\includegraphics{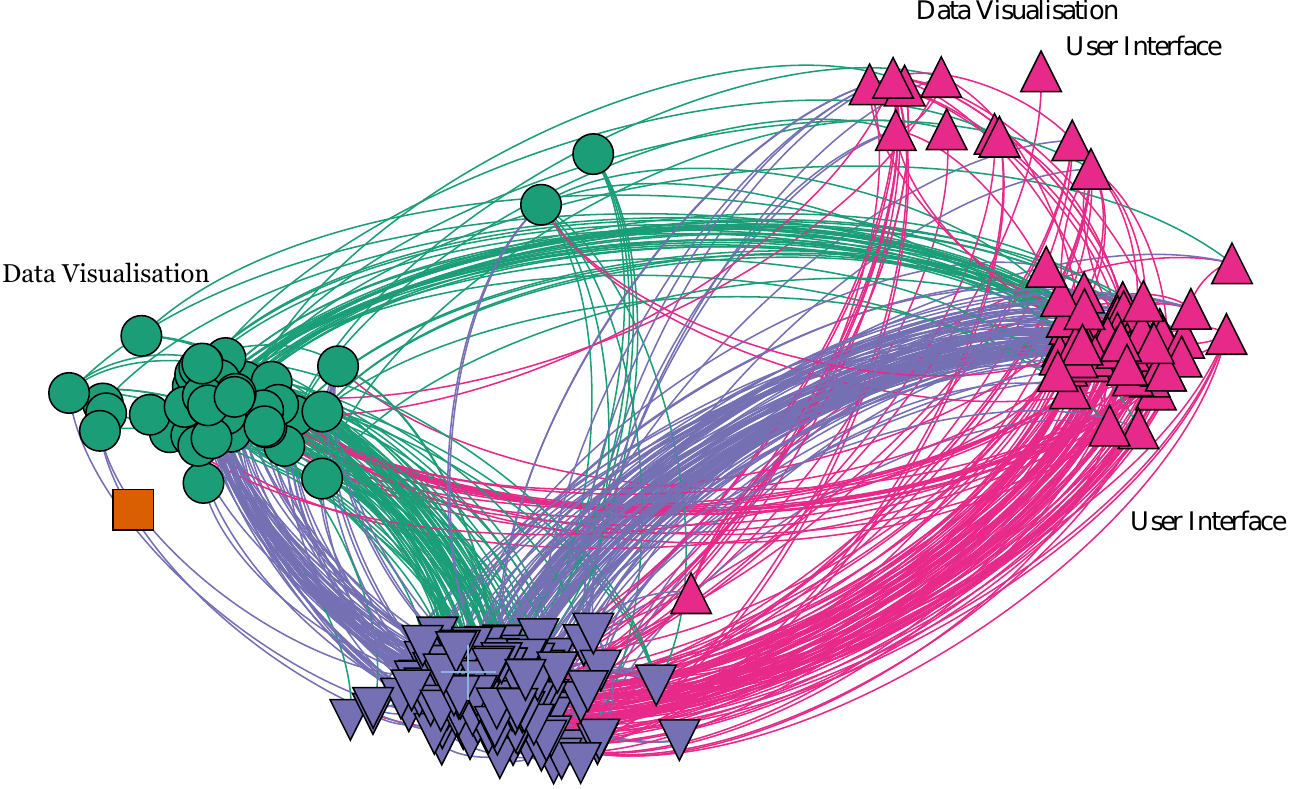}
	\caption{\bf The four cluster layout of the History of InfoVis dataset with graphTPP.}
	\label{fig10}
\end{figure*}

As per the \nameref{sec:inf} section above, initial exploration focuses on generating and analysing the optimum number of clusters followed by separation and analysis with graphTPP. The first layout generated has three main clusters  as shown in  Fig.~\ref{fig10} plus a fourth consisting of a single node.

The key attributes for this layout are the keywords user interface and data visualisation, previously described as burst words sparking in popularity between the years 1983--1991 and 1994--1995 respectively~\cite{Ke_2004}. In Fig. \ref{fig10} the nodes in the top right-hand corner have both  `Data Visualisation' and `User Interface' as significant attributes but some are assigned to the `User Interface' (UI) cluster while others to the `Data Visualisation' (DV) cluster.  There are fewer citations between these nodes in the mixed group to the DV cluster than between DV and UI but this may be an artefact of node volume. The two green circular nodes in the top centre of the graph have both the DV and UI attributes and cannot be moved towards either of the other groups through interaction. 

\begin{figure*}[tbp]
	\centering
	\includegraphics{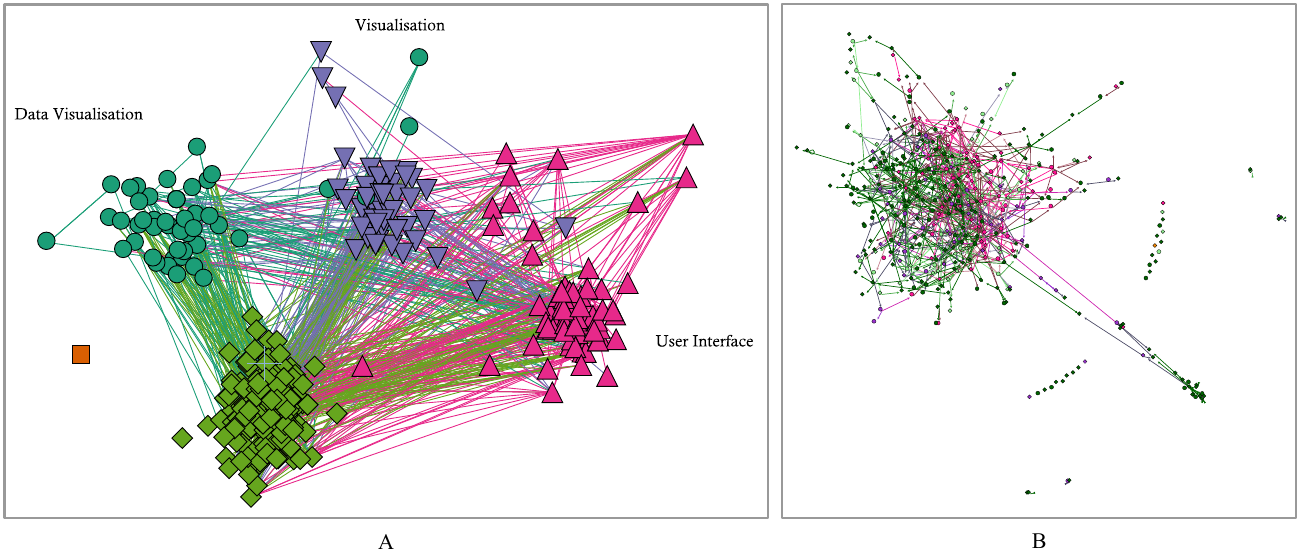}
	\caption{{\bf graphTPP vs a force-directed layout of the \emph{History of InfoVis} dataset.} The force-directed layout (B) uses the same colours as those given to the five clusters in the graphTPP layout (A).}
	\label{fig11}
\end{figure*}

The five cluster layout (Fig. \ref{fig11}(A)) shows a further attribute, visualisation (V), has been introduced as the main attribute of the fifth cluster. Comparing to a force-directed layout (Fig. \ref{fig11}(B)) shows that the force-directed algorithm is able to highlight a number of disconnected nodes as well as a small group of nodes ejected from the rest of the graph related to the keywords algorithm, animation, distribute, parallel and visualisation. This is the main interesting feature of this layout with the rest lacking any perceptible structure.

\subsubsection{Overlap between Visualisation and User Interface}\label{sec:citation_overlaps}

\begin{figure*}[htbp]

	\centering
	\includegraphics[width=\textwidth]{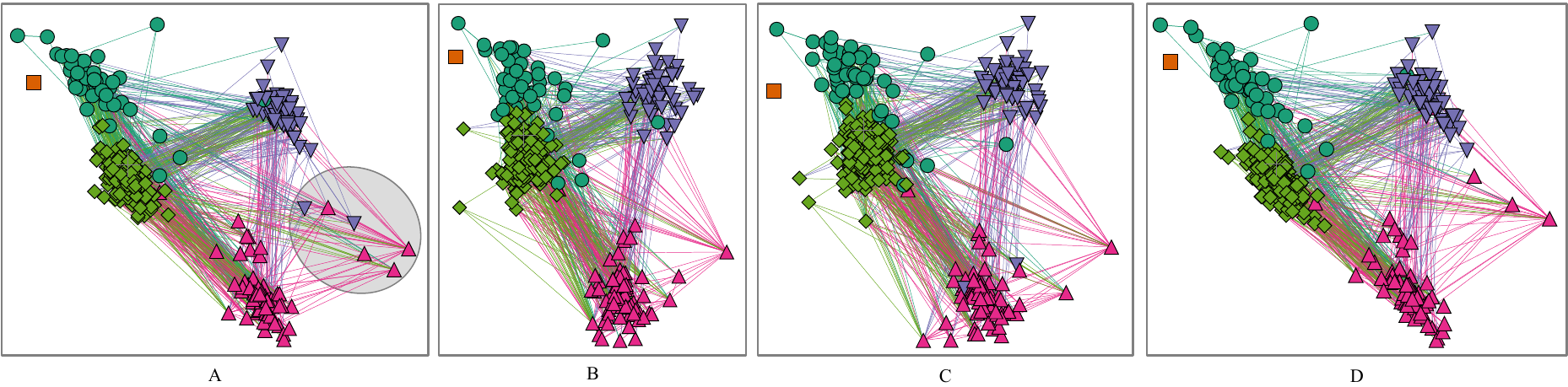}
	\caption{{\bf The four options for rearranging nodes in the graphTPP layout used in order to understand the overlap between the user interface and visualisation clusters.} (A) The original layout, (b) Nodes moved towards their own cluster, (C) Nodes moved towards the User Interface cluster, (D) Nodes moved towards the Visualisation cluster.} 
	\label{fig12}

\end{figure*}

In Fig. \ref{fig11} there are a number of nodes that sat between the clusters UI and V. Fig. \ref{fig12}(A) shows a layout for the five cluster graph with an overlap between the UI and V clusters. (A similar process to this can also be followed to investigate the overlap between the UI and  DV clusters.) There are three routes of investigation, starting from this projection, that could help to understand the relationships between these clusters: (1) move both sets of nodes back towards their `own' clusters; (2) move all the points to the UI cluster; or (3) move all the points to the V cluster.

The layouts in Fig. \ref{fig12} show the original layout (A) and the three resulting layouts if each of the possible routes of investigation is followed (B - D).  While some nodes move easily, others appear stuck. For example, when each set of nodes is moved towards its own cluster (Fig. \ref{fig12}(B)) all of the nodes except one move closer to their own group. The node that does not move is highly connected to the rest of the graph and cites many papers. This node represents Shneiderman's paper \emph{`The eyes have it: a task by data type taxonomy for information visualisations'}~\cite{Shneiderman_1996}.   Figs \ref{fig12}(C) and \ref{fig13}(D) show that there is not a projection where this node can be placed closer to the UI cluster group, hence it must be considered a true bridge between the areas. This is further supported by analysing the betweenness centralities in the network (number of shortest paths between nodes pairs that pass through that node) since it has the second highest betweenness centrality of all nodes in the graph. 

\subsubsection{A cluster-based edge-grouping method for improving graph clarity}\label{sec:citation_bundles}

An issue with the graphTPP layouts presented above is that as the graph increases in size the edge relationships become messy. This section presents an edge-grouping method to reduce the visual clutter in the graph and simultaneously improve the understanding of the between cluster relationships an is aimed at improving the overview layout rather than for investigating individual nodes. 

Edge bundling aims to reduce the visual clutter in network visualisations whilst still representing the high-level structure of the graph. Holten~\cite{Holten_2006} originally proposed the hierarchical edge bundling  which was followed up with a number of improvements including an adaptation to undirected graphs~\cite{Cui_2008,Holten_2009,Ersoy_2011,Gansner_2011,Luo_2012}.

The method proposed here is more primitive and aims to group the edges between each cluster by using the intrinsic properties of the layout to control the grouping. The edges are drawn by identifying the centre of each cluster, and then for each edge identify whether it is a intra-cluster or  inter-cluster edge. Intra cluster edges are drawn normally while inter-cluster edges are drawn as B\'{e}zier curves between the centroid of each cluster (offset slightly for each node to prevent overlapping edges and to create the impression of a thick line). 

\begin{figure*}[tbp]
	\centering
	\includegraphics{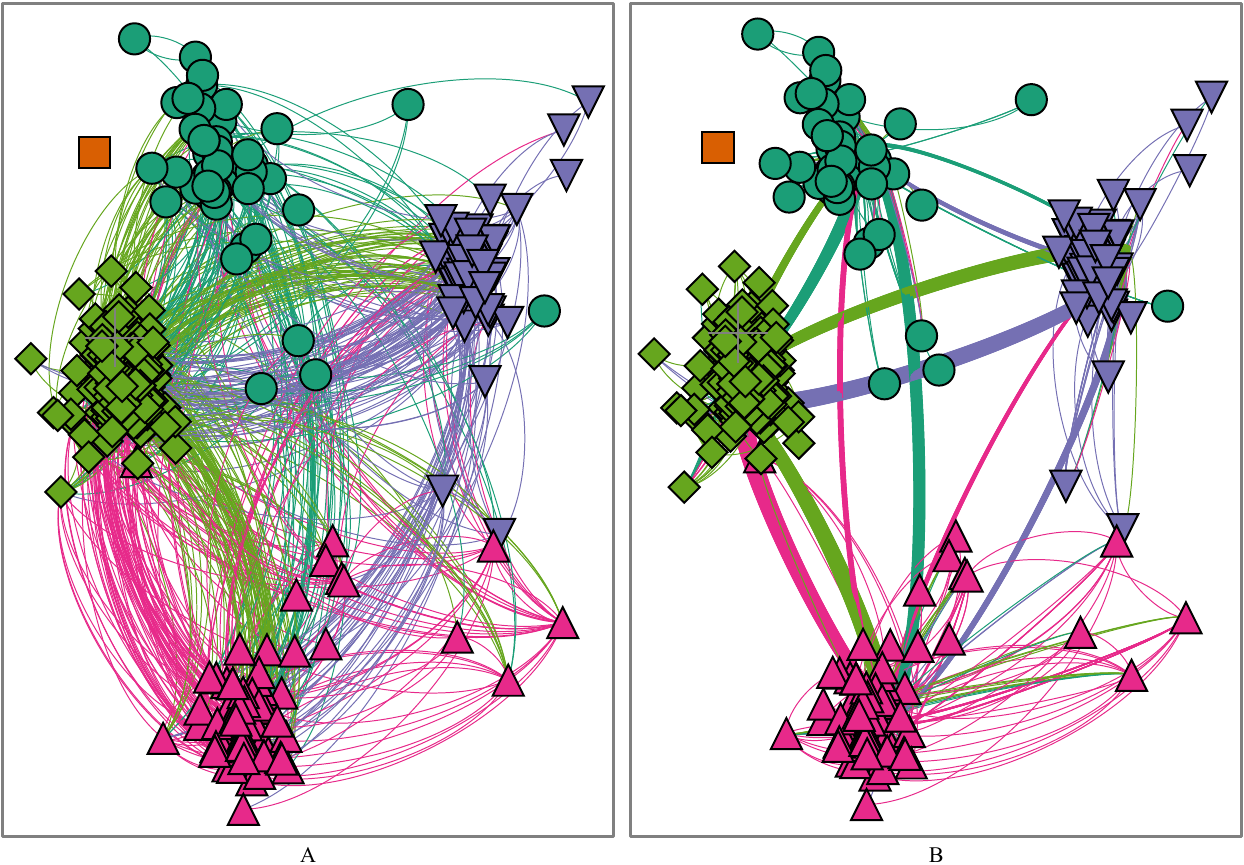}
	\caption{{\bf A comparison of the original five cluster layout with the grouped-edge layout.} (A) The original curved edges, (B) The edge-grouped layout}
	\label{fig13}
\end{figure*}

Fig. \ref{fig13} shows the difference between the original (A) and the edge-grouped layout (B) for the five cluster \emph{History of Infovis} graph. The grouped-edge layout presents a clear and  compact view of the edges where the thickness of the edge-groups is directly proportional to the total number of edges which enables inferences to be made about between-cluster influences. 

All clusters have the most edges to the `other' cluster, the second most edges to the UI cluster, third most edges to the V cluster and the fewest edges to the DV cluster. Between-cluster comparisons show that the UI group is more widely cited by both the V and DV groups than those groups are cited by research in the UI domain indicating a dependency from visualisation research  onto HCI type research. This may be attributed to timescales since UI was more established as a research field in 2004 than visualisation. 

\begin{figure*}[tb]
	\centering
	\includegraphics{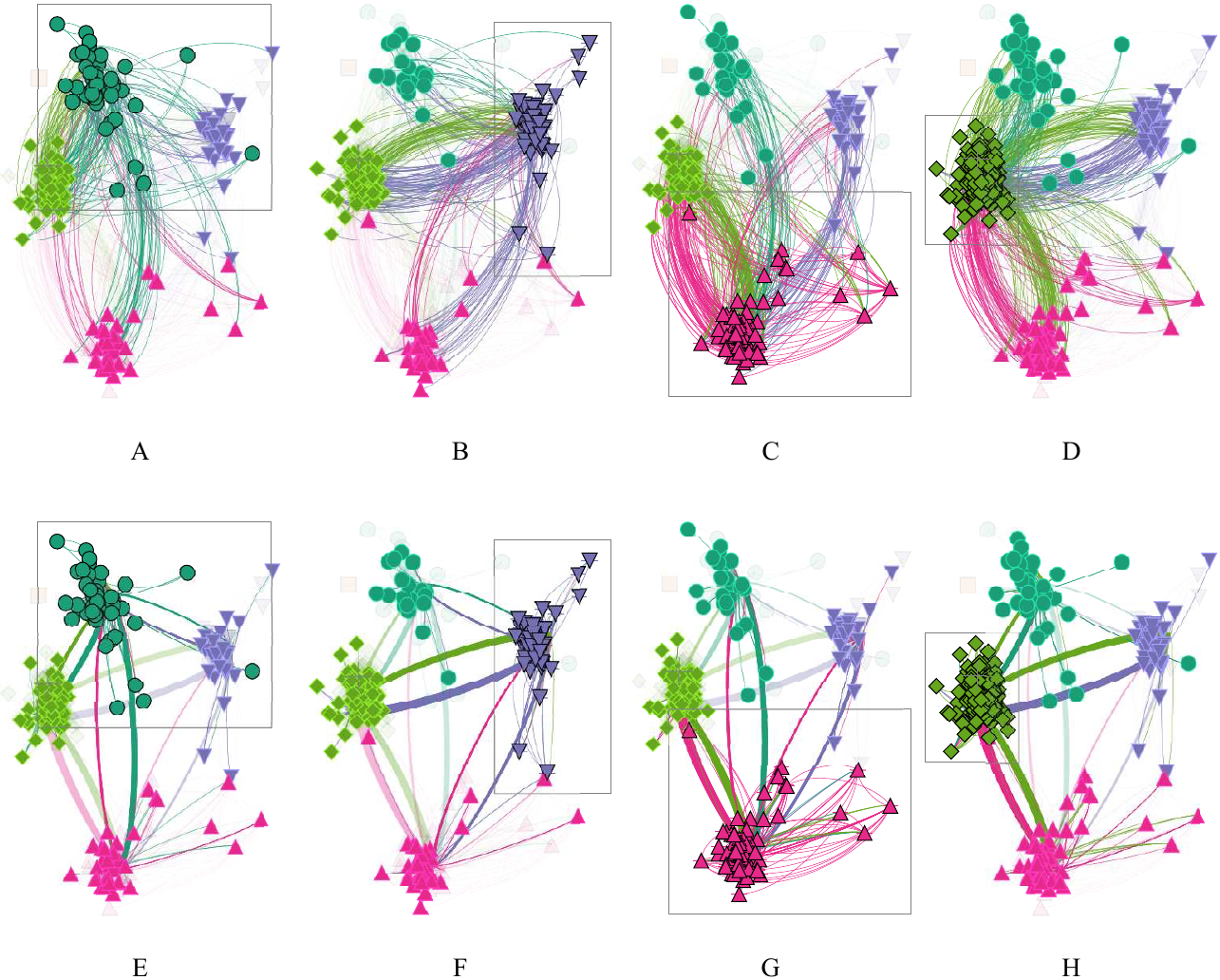}
	\caption{{\bf Comparing between-cluster edge analysis with curved and grouped edges using the five cluster graph with each of the clusters individually selected to focus on on only their incoming and outgoing edges in both the curved edge style and grouped edge style.} (A) Data Visualisation Curved Edges, (B) Visualisation Curved Edges, (C) User Interface Curved Edges, (D) Other Curved Edges, (E) Data Visualisation Grouped Edges, (F) Visualisation Grouped Edges, (G) User Interface Grouped Edges, (H) Other Grouped Edges.}
	\label{fig14}
\end{figure*}

The layouts in Fig. \ref{fig14} allow for comparisons between the curved edge and the grouped-edge approaches. The advantages of the bundled layout are clear: the reduction in clutter and the interpretable thickness of the line to determine the number of edges between the two clusters. Fewer individual lines also means that fewer nodes are obscured by the edges. A trade-off in producing a compact clustering is that the within-cluster citations are obscured, although, this is an issue for any edge style. The next section proposes the inclusion of edges as part of the projection. 

\subsubsection{Combining edges with attributes in the projection}\label{sec:citation_edgeatts}

\begin{figure*}[tb]

	\includegraphics{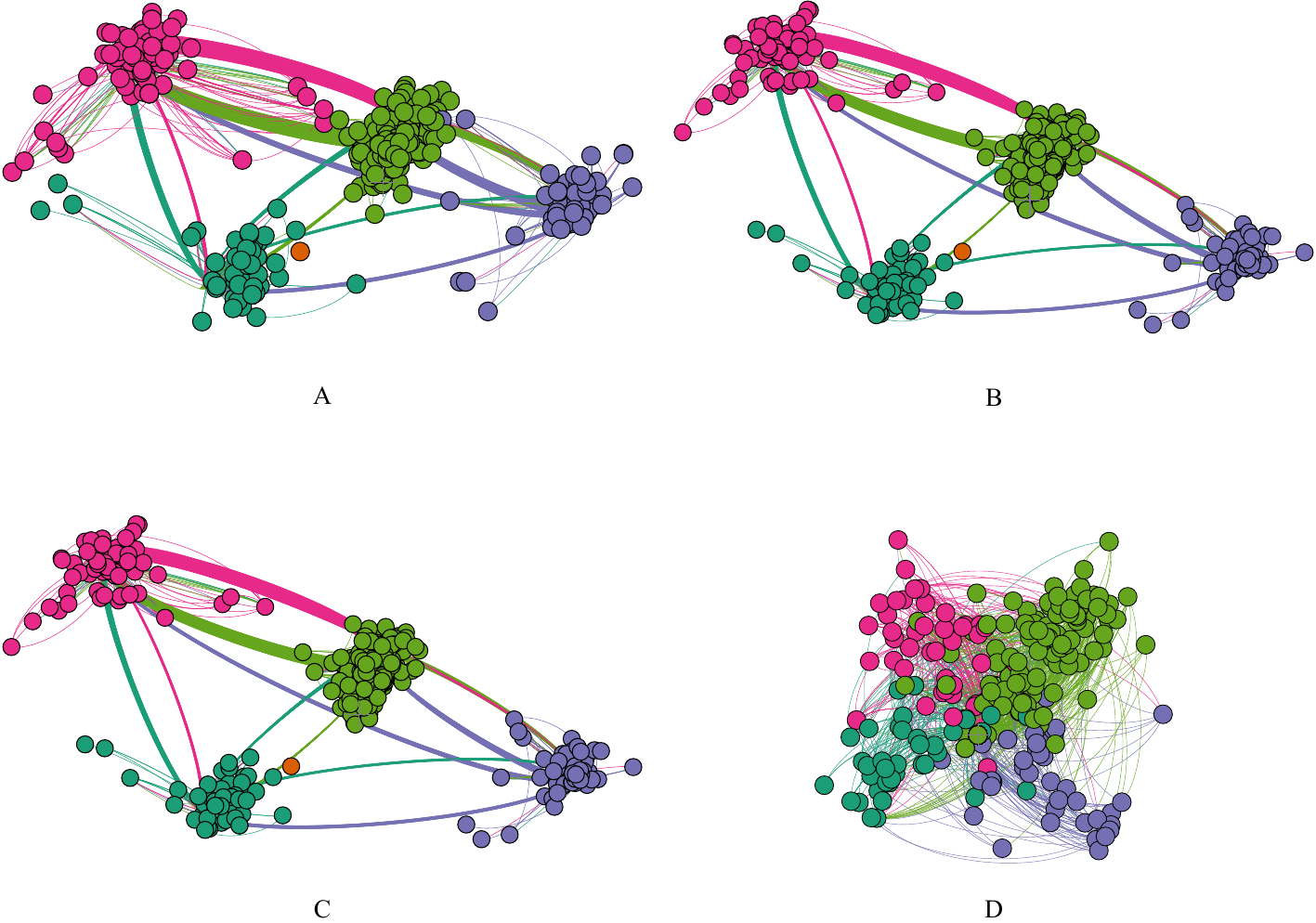}
	\caption{{\bf Four graphTPP layouts based on incorporating edges into the projection in the \emph{History of InfoVis} dataset.} (A) Original attribute only projection, (B) Attribute and Edges projection, (C) Attribute and Edges projection where in-degree was greater than one, (D) Edges only projection.}
	\label{fig15}

\end{figure*}

The central idea behind graphTPP has been that we need to rely on more than just the edge relationships to understand the graph: these relationships exist for a reason which we may be able to determine from a node's attributes. However, in a social network, for example, a friendship may form because a mutual friend introduces the parties to each other. Hence, it is this relationship that facilitates the formation of the friendship. In graphTPP this relationship is shown in the graph's edges but in larger graphs this may not be obvious or may be obscured. Thus an extension of graphTPP could be to include (some) relationships as additional attributes. Using the \emph{History of Infovis} dataset we tested this approach in three ways:

\begin{enumerate}
	\item Using the graph's adjacency matrix with each column as another attribute (Fig. \ref{fig15} (B)). 
	\item As (1) but only nodes with degree greater than one as attributes (Fig. \ref{fig15}(C)).
	\item Using only the adjacency matrix (but keeping the clusters generated by the original attributes) (Fig. \ref{fig15}(D)).
\end{enumerate}

Fig. \ref{fig15} shows the five-cluster graph laid out with the approaches detailed above. Fig. \ref{fig15}(A) shows the original keyword-only attribute-based projection as a reference. In Fig. \ref{fig15}(B), the attributes and adjacency matrix graph, each cluster is more compact and the overlapping cluster is less obvious. From this we can infer that nodes in the same cluster are more similar in their connectivity than nodes in different clusters. A caveat to these results is that in a linear projection each attribute has a weighting in the $x$ and $y$ directions. If an attribute is only associated with a single node these weightings can be manipulated meaning that  node can be positioned anywhere. When considering edges as attributes this occurs with all nodes that are connected to another node with degree one. This is acceptable for a few nodes in the graph but too much independence would mean that the layout no longer relies on the relationship to the attributes. 

The layout in Fig. \ref{fig15}(C) is a compromise between using all or none of the nodes as attributes. In this case, a node must have had an in-degree of greater than one, i.e., it must have been cited by more than one other paper to be included as an attribute thus ensuring there are no independent nodes. This threshold could be increased to reduce the number attributes if necessary; the layout is very similar to \ref{fig15}(B).

Fig. \ref{fig15}(D) shows the layout where we use the original cluster membership but the attributes are wholly defined through the adjacency matrix. This clarifies one point: there is still a lot of dependence in the layout otherwise the clusters would be easily separable. While nodes from each cluster are placed in the same area of the graph, there is not the clarity and separation seen in the other layouts. Ultimately, it shows some relationship between the graph's topology and the attribute generated clusters.

This has showed us that combining node-attributes with the adjacency matrix resulted in a better cluster separation. This may be because the nodes within a cluster share a more similar citation pattern than they do to nodes out-with their own cluster. This is especially useful for layout if a well-separated clustering is required but that the projection of the nodes should still be dependent on attributes. 

\subsubsection{Review of Bibliographic Visualisation}

The \emph{History of InfoVis} dataset included the citations and keywords of every paper presented at the InfoVis conferences from 1995 to 2003. The five cluster graph provided the most interesting layout features. Three clusters are  defined by the keywords Visualisation, Data Visualisation and User Interface. This does not preclude each node from having more than one of these as keywords, and, in fact, many do. This results in overlapping clusters in the layout and, rather than reducing the separation between the two clusters in the layout, the nodes which have both attributes are placed almost in a new cluster between them. Searching for a projection which allowed these nodes to be placed closer to their own cluster (i.e., the cluster that they were classified into using the $k$-means projection) found that while some nodes could be moved towards their own cluster successfully, Shneiderman's task-by-data-type-taxonomy article always stayed in a position that bridged the two areas: user interface and visualisation. 

The fact that the layout clearly shows these overlaps is an advantage of using graphTPP. A similar situation would occur if there was a subset of nodes within a cluster that differed in their attribute make-up. This provides an interesting starting point for exploration since while true exploration is a noble goal the user should also not feel at a loss as to how to proceed with the exploration and spotting an initial intriguing pattern is a good way to start. 
This section also introduced an edge-grouping method that reduces visual clutter while strengthening the overview view of the graph. It makes use of the layout's intrinsic clustering and the thickness of the group is directly proportional to the number of edges it represents. This is an advantage because at-a-glance interpretations of the connectivity between clusters can be made; however, some care needs to be taken in order to factor in the size of each cluster as more nodes are likely to result in more edges. The edge grouping is still more suited to overviews while the single line view is better for investigating individual edges or nodes. However, neither of these methods provides a suitable view of within-cluster edges.  

While edges within a cluster may not be immediately visible in the layout this does not mean that they cannot contribute to the it. This idea, to incorporate them into the projection, was explored above. Initially all the edges were used in the projection and hence every node became an attribute; then only those nodes with a in-degree greater than one (i.e., they were cited by more than one other paper) were used. There was little difference in the layouts for each of these cases but compared to the original attribute-only projection the effect of the edges on the projection became clear. The clusters were more compact and there were no overlaps between the user interface, visualisation and data visualisation clusters. This indicates there is more similarity between the connectivity patterns  of nodes in the same cluster than other clusters regardless of attributes. An edge-only projection showed that when using edges on their own a projection cannot be found that groups them into visually distinct clusters hence it is the combination of the attributes and edges that allows this. 

A limitation in this analysis of the History of InfoVis dataset is that the most recent time period it covers is already more than a decade ago and so the graph does not show more recent developments. Further, it is still difficult to convey the temporal nature of the publications and their citations which would also help to understand how the field developed. There is also the problem of how to filter the data. For example, after experimentation, a keyword had to be included more than five times to be considered as an attribute. This constraint meant that the nodes that did not possess any of the keywords selected could not be included in the projection since there would be nothing to base their projection position on. This means the remaining graph and layout presents an incomplete representation of the whole graph. 

Overall this section has demonstrated the use of graphTPP to explore citation networks by using a paper's keywords as attributes. The method has enabled the graph to be grouped into clusters to show the structural links between different topic areas and how they have developed. It particularly shows that much of the work in visualisation and data visualisation is underpinned by work in user interface design and that ideas flow between the two areas. It has also introduced an edge-grouping method and shown that using edges as attributes can contribute to effectiveness of the graphTPP layout. 

\subsection{Discussion of Case Studies}

The aim of the research described above was to find out whether  the use of multiple node attributes to interactively lay out graphs aid visualisation and analysis beyond using topological properties. To do this we have explored the use of graphTPP in three different case studies. Each case study has highlighted some advantages and disadvantages of using graphTPP as a layout method.

In the influence graph case study the layout and exploration process demonstrated graphTPP's capabilities as a tool for graph layout and analysis. It showed how the PCA view provided a good starting point for exploration and that cluster generation through $k$-means clustering could lead to interesting insights about the graph's structure that otherwise would not have been found. This emphasised how topological structure and attributes can go hand-in-hand to accentuate the understanding of the graph by showing relationships between clusters and their outliers. It also demonstrated that the user need not only rely on automatic cluster separation but directly manipulating the graph's layout themselves can lead to other insights.  graphTPP was again able to identify features of the graph that both EdgeMaps and the force layout were not. It also highlighted how in graphTPP's layout, the within-cluster edges are obscured as a consequence of the tight clustering. Relying on $k$-means clustering also means that the exploration of the network depends on the quality of the clusters generated. The graph is still small (226 nodes, 281 edges) and all the attributes are binary, although subsequent case studies demonstrated the use of graphTPP with graphs of larger sizes and numerical attributes. 

The graph produced for the network security case study has all the requirements for graphTPP (nodes, edges, numeric node attributes and known clusters). Its main strength was the at-a-glance visibility of outliers allowing for a quick decision on nodes that required further investigation but its most notable limitation was its inability to represent temporal data naturally (though graphTTP was not designed to do so). The layout of the firewall data was one attempt to overcome this limitation by defining attributes as the amount of activity per 15 minute period. However, in doing this, knowledge about which events contributed to the activity is lost and it did not prove to be a particularly effective solution. In terms of analysis the lack of domain knowledge was also a hindrance, although, this is not a limitation of the tool or method itself. In this case, graphTPP would be better suited to being part of a suite of analysis software for this type of data rather than acting independently.

The \emph{History of InfoVis} dataset demonstrated that graphTPP could show overlap between separate clusters and their analysis in the context of the graph such as Shneidermans's paper that bridged the two areas of user interface and data visualisation. The section also showed how edge grouping could result in a stronger visual overview of the clusters, and that including edges in the projection produced a more compact clustering. This furthers the relationship between the attributes, the graph's topological structure and the layout. A particular issue was that these networks suffered from incomplete data which resulted in the exclusion of several papers; however, it is unlikely that all the data is always available in any situation. The aim of the \emph{History of InfoVis} dataset was to track the development of this area over time but graphTPP was not able to fully capture the temporal relations. Regardless, prior knowledge associated with the areas of HCI and user interface meant that the graph gave an impression of how certain areas had evolved. This was an advantage gained through familiarity with this area of research and it could be expected that if this method was applied to other domain areas then researchers in those fields would be able to gain similar insights. 
		
\section{Lessons learned and recommendations for future work}\label{sec:ll}
In carrying out this research we have identified a number of open questions that are still to be solved within graph layout and particularly for multivariate graphs. These lessons learned have also then led to a number of avenues for future work. 

\subsection{The future of (multivariate) graph layout}

\subsubsection{What is the best way to draw a graph?}

Blythe \emph{et al.}'s~\cite{Blythe_1996} statement: ``the best drawing is the one that highlights the characteristic of the network that is being discussed'' sums up the dilemma that many graph visualisations face. A layout that shows too much, too little or even has the wrong emphasis may conceal important  features of the graph. Finding the `best' layout is a delicate balance of a number of factors including: 

\begin{itemize}
\itemsep 0em
	\item graph size and edge density,
	\item data type and number of attributes available,
	\item the balance between presentation and exploration,
	\item prior domain knowledge,
	\item the purpose of the layout,
	\item which level the analysis take place: macro, relationship, micro,
	\item amount of aggregation tolerated, 
	\item the familiarity of the metaphor (to the user) --- node-link, matrix, etc. 
\end{itemize}

Understanding these factors will assist in determining a suitable layout method appropriate to the dataset and the tasks to be accomplished.

Our work in the above \nameref{sec:cs} has established that graphTPP is most effective on medium (up to 1000 nodes) sized node-attribute graphs composed of numeric attributes and one or more attributes that describe cluster membership. The goal is exploratory and takes place at the overview and relationship levels. The user's principal task is to produce a clustered layout that facilitates the investigation each cluster's make-up. The aim is to hypothesise how topological structure,  attribute and clusters are related together with notable outliers. 

Thus, the most important steps in determining which is the best way to draw a graph is to determine which types of data are to be represented by the layout and what questions the layout should help to answer. A consequence of this idea would be  a process that prescribes particular layouts to suit specific graph structures, attribute data and specific tasks. 

\subsubsection{What is the role for attributes in layout?}
The contribution of attributes in graph layout is still an important consideration. In application areas such as biology and social sciences graphs are often supplemented by node and edge attributes. Understanding the dependencies between attributes and a graph's structure is essential for eliciting knowledge about the graph. Furthermore, the order in which these dependencies are considered is also significant. Typically a topological layout is produced and attributes are visually encoded onto the nodes and edges; however, this only answers the question ---  given the topology of the graph, how are the attributes related to it or, at least to the topology implied by the layout technique used? This work has focused its attention on the reverse of that question: given the  similarities between attributes, how does the topology of the graph relate to these attributes? Both questions are worth asking and may elicit different results. Consequently, it is important to be aware of the limitations and accuracy of the layout. Similarly, in the attribute case, the processes of how that similarity is represented in 2D space is also essential. 

Hybrid techniques consider attributes and topology concurrently. Most interestingly, van den Elzen and van Wijk~\cite{elzen_2014} present a system for concurrent attribute and topology exploration that provides additional visualisation of attributes and also an aggregate high-level overview. However, layout is restricted to using only two attribute dimensions or a force-based layout. In MagnetViz~\cite{Spritzer_2011}  users select attributes to control the forces in the layout but only a few attributes can be used at a time and the user already needs a good knowledge of them. This is in contrast to graphTPP where interesting attributes emerge from the layout.  

The attribute's type (numeric, nominal, binary, set, temporal, etc.) also influences how it can be represented. Certain types of attributes will be more suited to particular attribute layout methods; for example, graphTPP used numeric or tags and sets converted to binary data for describing dimensions while nominal data was used to define the clusters. Further, graph metrics such as centralities may also be incorporated. The domain area may also impose specific requirements on the layout thus representing the graph and its attributes in a way that respects these requirements is also essential. 

\subsubsection{What are the alternative ways of representing attributes?}
Attributes are often visually encoded onto the nodes or edges as colour, size, position, orientation, shape and texture. Additional attempts to include more information on the nodes has been to represent them as glyphs or even pie charts but these become difficult to interpret. graphTPP and other dimension reduction graph layout techniques represent attributes by position and  proximity to other nodes. The force feature-space data-transform~\cite{Veljkovic_2008} improves class separation when using dimension reduction techniques. However, graphTPP has shown class separation comes at the expense of the view of the within cluster edges and its effect on the interpretation of the projection would also have to be monitored. 

Barsky \emph{et al.}'s~\cite{Barsky_2008} use of hierarchy to represent cell structure may  provide inspiration on how attributes related to real-world phenomena can be used. They also incorporated a parallel coordinate plot of attributes  with the graph to provide additional context and the inclusion of further coordinated visualisations  would be another method of analysing attributes without cluttering the graph visualisation (e.g.~\cite{elzen_2014}). 

\subsubsection{How to assess the strength of the link between attributes and the structure of the graph?}

In some domains there is a strong association between the attributes and the topological structure of the graph. In social networks communities form based on language, age and race~\cite{Newman_2003}. These mixing patterns, known as assortative/disassortative are commonly used in the network science literature but visualisations explicitly exploiting their features are less prevalent. Opportunities may include exploration of whether a mixing pattern exists, to determine which attributes are strongly associated with the topology, or, to validate models-based  patterns identified in existing layouts which could then be automatically applied to a new graph for layout or lead to further exploration if the model does not produce the expected layout.

The ultimate aim would be to have a set of layouts that are known to be suited to particular graph features and can be determined before the layout process commences. This  may include specific layouts that depend both on structural features and on the domain area.

\subsubsection{Clustering as a structural layout feature}

A key reason for selecting TPP was its bias towards clustering. This was motivated by van Ham and van Wijk~\cite{vanHam_2004} who found that manual user layouts often displayed a preference for clustering. Understanding the relationships between clusters then became one of the principal tasks associated with the graphTPP layout. A limitation of the use of clustering is that a cluster is perceived as having high interconnectivity but this may not be the case in practice. Thus, it has to be clear what a cluster represents: in graphTPP it is only attributes' similarity. Any layout that tries to optimise for clustering will have issues with within cluster edges being obscured unless they develop a specific representation to counteract it. 

McGrath, Blythe and Krackhardt~\cite{McGrath_1996} warned layout designers to be careful about cluster presentation and its implications on user perceptions of the graph. They found that the same graph with three different layouts altered the number of user reported clusters and that there may be no within-cluster edges. The advantage of systems such as NodeTrix~\cite{Henry_2007} is that the matrices give a very clear view of the sparsity or density of each cluster but this is only effective up to a limited cluster size. 

The decision is always between imposing a clustered structure on the layout, whether to see if one emerges naturally or to lay out the graph first and then measure how well that fits in with the prior expectations of clustering. This is a question for those who are producing the layouts and depends on the type of task they want to use the layout for. However, those who produce layout algorithms also have a responsibility to inform or educate users about the contexts of how that layout should be perceived.

\subsubsection{Is dimension reduction a good method for graph layout and  analysis?}

This research is grounded in the idea that dimension reduction can be used effectively for finding an interpretable representation of a graph. Dimension reduction can be used on graph-theoretic distances with MDS  to produce a layout or node-attributes can be used to determine the dimensions and methods such as graphTPP and others can be applied. Since there are so many dimension reduction techniques available there should be plenty of options to find an algorithm that suits the data available. However, this makes it important to select one which is compatible with the dataset and that the results are interpreted correctly. For example, MDS produces reasonable layouts using graph-theoretic distances for similarity but inputting graph-theoretic distances as attributes in graphTPP was unable to produce any acceptable layout. Chuang \emph{et al.}~\cite{Chuang_2012} also noticed that supposedly interesting features of the projection can sometimes just turn out to be artefacts.  The interaction in graphTPP is one solution to counteract this as by attempting to move one set of nodes the user can see if the other nodes that they thought were similar will follow. 

No matter the algorithm, a potential problem of using a dimension reduction algorithm is that the nodes are not area-aware. This results in many overlapping nodes which is not always a desired feature and can become a major problem if much of the graph is occluded, although a post-processing solution could still be applied. 

\subsubsection{What is the future for node-link diagrams?}

Node-link diagrams are a popular visual metaphor for representing a graph as they intuitively represent the concepts of a graph. That is, a graph describes the relationships between a number of entities and a node-link diagram concretely shows this by means of drawing a line between two other objects.
However, this does not naturally imply that it is also the most effective representation. Matrix representations of graphs (each row and column is a node and an edge is usually shown by a filled square at the intersection of source node row and the target node column) have been shown to be more successful at completing a variety of generic graph based tasks than node-link diagrams except for path-finding~\cite{Ghoniem_2005}. MatrixExplorer~\cite{Henry_2006}, NodeTrix~\cite{Henry_2007}, GraphPrism~\cite{Kairam_2012} and BioFabric~\cite{Longabaugh_2012} all use matrix-based metaphors to display graphs. One of the challenges in matrix representations is in the node-ordering and there are few options for the inclusion of attributes. 

For node-link diagrams, their most popular representation are force-directed layouts. One option to consider would be whether force layouts could be extended to take into account attributes and other structural features. For example, are there ways to advance on the ideas of MagentViz by incorporating more of the attributes in the attraction/repulsion calculation or methods that can encourage the layout to highlight structural features more clearly such as nodes with high degree, centrality values, bridges between clusters, or  similar connectivity patterns? 

It is likely that the node-link representation will remain popular. However, continuing to draw larger and larger graphs, whether hairball in appearance or not will, if they are not already doing so, push the limits of both technology and users' perceptual abilities. Thus, not only is it important to think about the future for node-link diagrams and how they can become `smarter' it is also necessary to think about pre-processing and how much of the graph  to visualise at any one time, and when aggregation should be used.

\subsubsection{Can we always show `everything'?}

An issue with any visualisation, where the screen size and resolution are limited, is how much of the data should be shown in any one view at any one time. That is, at which point should data aggregation happen and why? In the case of graphs it is, perhaps, one of the most difficult decisions when designing a visualisation since every connection in the network can be seen as vital and there is no clear way to determine which may be significant and which may be irrelevant. One of the sources of confusion for showing `everything' seems to be that showing everything equates to a good overview of the graph. One only has to observe hairball style visualisations to see that showing everything sometimes only tells you nothing, or even obscures the interesting information. Thus, the real questions should be what is a good overview and how do we produce one without showing everything?

Aggregation is one method for simplifying the graph either computationally or visually.  In a graph, selecting how the aggregation should occur is the first step. Either or both the nodes and edges can be aggregated and this can be based on their topological structure, such as uniting nodes with similar connection patterns or using glyphs to represent them~\cite{Dunne_2013}, through interactive aggregation~\cite{Lin_2013}, or using attribute similarity. Edge-bundling is a popular method of aggregating edges and this can efficiently reduce clutter and expose edge routing patterns previously hidden behind the hairball.

\subsection{Avenues for future work}

Through the course of this research a number of areas for possible future work, both based on graphTPP and in the wider multivariate graph visualisation community, have been identified.

\begin{itemize}

	\item graphTPP enabled the separation of clusters into distinct visual areas in order to analyse between-cluster behavior and outliers while within-cluster edges were often obscured. The next step would be to develop a scalable solution to visualise within-cluster edges in graphTPP or any other clustered layout. 

	\item Over a certain size, graphs visualised with graphTPP begin to clutter the display, hindering analysis. Aggregating nodes within a cluster may help ease some of the complexity. Further, aggregation for all graph layout methods may be considered. 

	\item In the \nameref{sec:bib} section above graphTPP showed that Shneiderman's task-by-data-type-taxonomy paper bridged two areas and that it was subsequently found to have high betweenness centrality. Layout methods that bring out important structural features of the graph should be considered. 

	\item The motivating ideas for the graphTPP layout method are closely aligned with assortative mixing patterns. Incorporating more theories from network science into goals for network layout is another promising research direction. 

	\item For most graph layouts the topological structure of the graph is used to lay out the graph first, then the attributes are incorporated. In graphTPP the attributes were used to position the nodes first and then the graph was added. Further work on comparing how these different perspectives affect viewers' perception and understanding of the graph is required. 

	\item graphTPP is based on linear projections. The visualisation is, in effect, the product of the multiplication of two matrices: the original data and the linear projection. Given two datasets with the same attributes but different values, the application of the linear projection matrix from one to the other would test the similarity between the two datasets. In time, this could lead to more prescriptive layouts. 

	\item graphTPP required a graph structure, numerical attributes and, preferably, at least one known cluster as a minimum for layout. Other layouts also have optimum conditions under which they may produce their best results. Tailoring layouts to specific graph structures, domain areas, attribute types and also tasks would aid users when selecting the layout that they wish to use. 

	\item Much work has gone into developing force-directed methods but they are not always the best solution for graph layout. Methods such as LinLog and OpenOrd re-purposed aspects of force layouts to highlight clustering while MagnetViz and GraphScape did it to include attributes. There must be further structural features that force layouts can be manipulated to show or ways in which attributes can be incorporated along with possible aggregation schemes. 

	\item Task is a very important consideration in network visualisation~\cite{pretorius2014}. How layout affects task completion on graphs larger than 20-30 nodes is not well studied. Neither is the question of which tasks are best suited to which type of layout and what tasks can be carried out on which sizes of graph. 

	\item The section on \nameref{sec:netsec} discussed how additional visualisations may have supported the analysis in the network security task. The incorporation of other linked visualisations to improve the understanding of attribute distribution and other features of the graph would provide a more integrated approach to analysis. 
	
\end{itemize}

\section{Conclusions}\label{sec:conc}

In this paper we have presented graphTPP as a method to drive the graph layout of node-attribute graphs via the dimension reduction method TPP in order to produce a layout that gives further insight into the dependencies between the graph's topological structure and its attributes. This was motivated by the idea that nodes with similar attributes are also more likely to be related. We have demonstrated, through three diverse case studies, that graphTPP is able to elicit insights into these relationships that other layout methods currently fail to provide. graphTPP is able to support a strong visual separation of nodes based entirely on the value of their attributes, a clear indication of where outliers appear, and, importantly, allows users to understand which attributes contribute to a particular clustering. The layout supports interaction through direct manipulation of the nodes facilitating the user in hypothesis formation and testing. Additionally, we presented an extension to the layout that produced a grouped-edge effect for a clearer overview layout and explored the use of attribute and adjacency data combined to drive the layout. 

This research demonstrated that node-attributes have an important role to play in layout but also identified a number of limitations that apply specifically to graphTPP and to network visualisation as a whole. In particular this included the ability to show relationships within dense clusters, restrictions around the size of the graph which can be visualised without aggregation, the requirement for additional layouts to support temporal attributes and the need to incorporate more theory from network science into layout.

\section{Acknowledgments}
This research was funded by Northumbria University.

\bibliographystyle{IEEEtranS}

\end{document}